\documentclass[aps,pre,twocolumn,superscriptaddress,floatfix,longbibliography,nofootinbib]{revtex4-2}
\usepackage[T1]{fontenc}
\usepackage[utf8]{inputenc}
\usepackage{newtxtext}
\usepackage{newtxmath}
\usepackage{amsmath}
\usepackage{bm}
\usepackage{graphicx}
\usepackage{dcolumn}
\usepackage{booktabs}
\usepackage{array}
\usepackage{xcolor}
\usepackage{microtype}
\definecolor{linkblue}{HTML}{2A3E78}
\usepackage[colorlinks=true,linkcolor=linkblue,citecolor=linkblue,urlcolor=linkblue]{hyperref}

\newcommand{\lmin}{\ell_{\min}}
\newcommand{\lmax}{\ell_{\max}}
\newcommand{\Lang}{\mathcal{L}}
\newcommand{\mub}{\mu_{\rm bare}}
\newcommand{\Jup}{J_{\rm up}}
\newcommand{\Jcond}{J_{\rm cond}}
\newcommand{\E}{\mathbb{E}}
\newcommand{\figref}[1]{Fig.~\ref{#1}}

\newif\ifrevision \revisionfalse
\definecolor{revred}{HTML}{C00000}
\DeclareRobustCommand{\rev}[1]{\ifrevision{\color{revred}#1}\else#1\fi}

\begin{document}

\title{Memory in biological L\'evy walks from the dynamic-entropy formalism}

\author{Adriano Barra}
\affiliation{Istituto Nazionale d'Alta Matematica, GNFM, Rome, Italy}
\affiliation{CNR-Nanotec, Unit\`a di Lecce, Lecce, Italy}
\affiliation{Dipartimento di Scienze di Base e Applicate all'Ingegneria, Sapienza Universit\`a di Roma, Rome, Italy}
\affiliation{Istituto Nazionale di Fisica Nucleare, Sezione di Roma 1, Rome, Italy}
\author{Andrea Ladiana}
\email[Corresponding author: ]{andrea.ladiana@uniroma1.it}
\affiliation{Istituto Nazionale d'Alta Matematica, GNFM, Rome, Italy}
\affiliation{Dipartimento di Scienze di Base e Applicate all'Ingegneria, Sapienza Universit\`a di Roma, Rome, Italy}

\date{\today}

\begin{abstract}
     \rev{L\'evy walks are recognized by the histogram of their flight lengths, while the renewal hypothesis on which the textbook model rests, that every flight is drawn afresh, is seldom tested. We test it with the dynamic-entropy formalism. Maximizing the causal entropy of a sequence of logarithmic flight lengths at fixed lagged correlations yields a chain of continuous spins on a compact interval, whose depth, selected by the data, tells a renewal walker from a Markovian one and from one with a longer memory. With one lag the chain is exactly solvable. A duality that exchanges long and short flights locates a first-order condensation of the flights at the upper cutoff, with the logarithmic range of the flights in the role of system size, and a positive memory makes the histogram heavier-tailed than the rule that generates each flight. For groups, the same construction separates the memory of each individual from delayed interactions and from drives common to all units, which an analysis of the group average confounds, and it gives the collective speed a critical point. On open data, renewal is rejected in seven of ten conditions of \emph{Drosophila} larvae and in every termite, whose memory reaches a few flights back and lowers the exponent of the histogram by more than half a unit below that of the one-step rule. Wild baboons align with where their nearest neighbors, those ahead above all, were heading 10 to 30\,s earlier, and homing pigeons are coupled across their small flocks. Zebrafish T cells, a negative control, remember their own speed, but no delayed coupling between the speeds of different cells survives the controls.}
\end{abstract}

\maketitle

\section{Introduction}\label{sec:intro}

Put a \emph{Drosophila} larva on a dish of plain agar, with nothing to smell, see or eat, and it will not stay put. It crawls in straight runs separated by turns: many short runs and, every now and then, a very long one, with lengths spread over two decades in a way that resembles a power law~\cite{Sims2019}. Termite workers in a Petri dish~\cite{Castiblanco2022}, T cells in living tissue~\cite{Jerison2020} and marine predators in the open ocean~\cite{Humphries2010} leave records of the same kind. A walk made of straight flights whose lengths $\ell$ follow $p(\ell)\propto\ell^{-\mu}$, with $1<\mu\le3$, is called a L\'evy walk~\cite{Viswanathan1999,Zaburdaev2015}, and whether real organisms perform L\'evy walks, and why, has been argued about for two decades~\cite{Edwards2007,Pyke2015,Reynolds2018}.

Most of that argument concerns the histogram of flight lengths. This paper is about their order. Does a long flight make the next one longer? Does a walker remember its previous flights, and if so, how many? And if it does, is the exponent read from the histogram still the exponent of the rule that the walker follows at each turn? The textbook L\'evy walk has no memory: it is a renewal process, in which every flight is drawn afresh from the same law. A power law, however, only counts long flights. It does not say whether they come in runs, and runs of long flights change how far and how fast a walker spreads~\cite{Aghion2021}. When walkers move in groups a second question joins the first: when a walker speeds up or turns, is it following its own past or the past of the others?

Questions of this kind are inverse problems, and maximum entropy (MaxEnt) is a natural way to pose them~\cite{Jaynes1957}. Among all the probability laws that reproduce a chosen set of measured averages, MaxEnt selects the one that assumes nothing else. When the averages are those of some functions $f_k$, the selected law is an exponential family, $p\propto\exp(\sum_k\lambda_kf_k)$, formally a Boltzmann weight whose Lagrange multipliers $\lambda_k$ act as fields and couplings; read in this way, the correlations measured in flocks of birds and in populations of neurons have been turned into networks of effective interactions~\cite{Bialek2012,Mora2016,Roudi2011,Nguyen2017}. Maximizing the entropy of whole trajectories rather than of single configurations, the principle of maximum caliber~\cite{Presse2013,Ge2012}, carries the idea over to dynamics. Constraints on correlations at different times have been used to predict the spiking of neurons~\cite{Marre2009} and the motion of birds in a flock~\cite{Cavagna2014}. In cell migration, inference of this kind has told collective from individual motion~\cite{Agliari2020}, measured how a drug reshapes the interactions between cancer and stromal cells~\cite{Alemanno2023} and, in the dynamic-entropy formalism of Agliari \emph{et al.}~\cite{Agliari2024}, inferred delayed interactions between migrating cells from the directions in which they move.

We apply this program to the lengths of flights. Its static half is old: fixing only the mean of $\ln\ell$, MaxEnt returns a power law~\cite{Montroll1983}. The dynamic half is not. Fixing also the correlations between $\ln\ell_t$ and $\ln\ell_{t-\tau}$, for lags up to a depth $K$, turns the sequence of flights into a chain of spins coupled in time, and the depth that the data select tells a renewal walker ($K=0$) from a Markovian one ($K=1$) and from one with a longer memory ($K>1$). The chain is a small statistical-mechanics problem of its own. With one lag it can be solved exactly. Exchanging long and short flights, $\ell\to\lmin\lmax/\ell$, maps the chain onto itself with different parameters, and on the line of parameters that this duality leaves unchanged the flights condense at the largest available scale through a first-order transition, with the logarithmic range of the flights in the role of system size. Short of the transition, memory makes the histogram heavier-tailed than the rule that generates it. In a group, the same construction separates the correlation of a walker with the past into a part due to its own past and a part due to the past of the others, and the collective speed acquires a critical point.

On open data, memory turns out to be common. \emph{Drosophila} larvae crawling without sensory cues~\cite{Sims2019} remember one to four flights back, and they remember most when their brain lobes and sensory neurons have been ablated. Termites~\cite{Castiblanco2022} remember about four flights back, and the wall of their dish, which they follow closely, does not explain it. In groups, wild baboons~\cite{StrandburgPeshkin2015} align with where their nearest neighbors, those ahead above all, were heading 10 to 30\,s earlier, and homing pigeons~\cite{MehlhausenFranks2025} are coupled to the whole of their small flocks. In zebrafish T cells~\cite{Jerison2020}, which we use as a negative control, the delayed coupling that an analysis of their group average would report is the memory of each cell, and no coupling between the speeds of different cells survives the controls; this does not mean that T cells do not interact, only that their interactions do not take this form (Sec.~\ref{sec:tcells}).

Rather than evaluating search optimality, competing functional distributions (such as log-normal alternatives over empirical ranges; Appendix~\ref{app:data}), or underlying neural circuitry, our objective is to characterize trajectory dependence directly: determining whether sequential flight memory exists, quantifying its range, and isolating individual persistence from social coupling.

Section~\ref{sec:model} builds the model, for one walker and for many, and Sec.~\ref{sec:statmech} works out its emergent properties. Section~\ref{sec:inference} turns the model into an estimator and Sec.~\ref{sec:data} applies it to larvae, termites, baboons, pigeons and T cells. Section~\ref{sec:discussion} discusses what the results mean for L\'evy walks. Proofs, validations on synthetic data and the controls on the real ones are collected in the Appendices.

\section{Flights with memory}\label{sec:model}

\begin{figure*}[t]
\centering
\includegraphics[width=\textwidth]{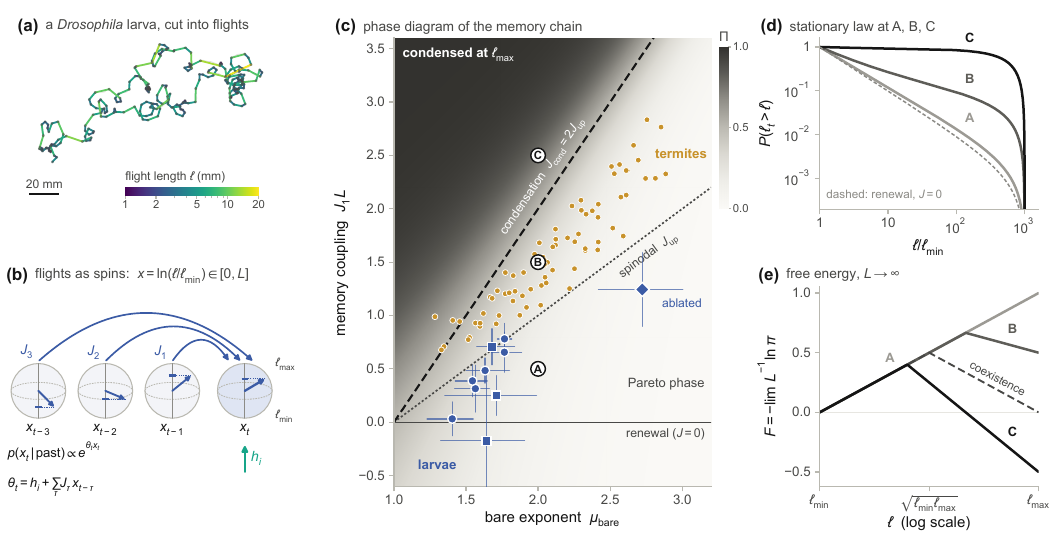}
\caption{From a track to a chain of spins. (a)~One hour of a \emph{Drosophila} larva (shi/+, 33\,$^\circ$C) crawling on plain agar: the flights (black segments, dots at the turns) cut out of the recorded track (grey). (b)~In the variable $x=\ln(\ell/\lmin)\in[0,L]$ a flight is the vertical projection of a classical spin on a sphere whose poles are the cutoffs $\lmin$ and $\lmax$; shown are four consecutive flights of the larva in (a). The next flight $x_t$ feels a field $h_i$, specific to the walker, and the previous flights through the couplings $J_\tau$, Eq.~\eqref{eq:causal}. (c)~Phase diagram of the chain with one lag ($K=1$). Horizontal axis: the bare exponent $\mub=1-h$, with which a flight is drawn after a run of the shortest ones (Sec.~\ref{sec:causal}). Vertical axis: the coupling $J_1$ between consecutive flights times the logarithmic range $L$, the largest change of exponent that one flight can induce in the next (Sec.~\ref{sec:path}). Shading: fraction $\Pi$ of flights longer than the geometric mean $\sqrt{\lmin\lmax}$, the midpoint of the logarithmic range, which stays fixed when long and short flights are exchanged; it is computed from the exact stationary law of the chain, Eq.~\eqref{eq:order}, for $\lmax/\lmin=10^2$. Dotted, the coupling $\Jup$ beyond which a run of long flights can sustain itself (spinodal); dashed, the coupling $\Jcond=2\Jup$ beyond which the flights condense at $\lmax$, Eq.~\eqref{eq:thresholds} (Sec.~\ref{sec:condensation}). Indigo: larval conditions (median and interquartile range of $\mub$ over larvae, $\pm2$ standard errors of $J_1L$; circles, controls; squares, neurons silenced; diamond, brain lobes and sensory neurons ablated); ochre: 75 termite workers (Sec.~\ref{sec:data})\rev{; both from fits with $K=1$}. (d)~Stationary distribution of flight lengths at the three states marked A, B and C in (c), which share $\mub=2$ and have $J_1L=0.5$, 1.5 and 2.5, for $\lmax/\lmin=10^3$; dashed, the renewal walk with the same $\mub$. (e)~Free energy $F(u)=-\lim_{L\to\infty}L^{-1}\ln\pi$ along $u=\ln(\ell/\lmin)/L$, Eq.~\eqref{eq:freeenergy}. It is piecewise linear because the stationary weight is a product of exponentials in $x$. In A the only minimum is at $\lmin$; in B a second, higher minimum at $\lmax$ makes runs of long flights metastable; on the dashed coexistence line the two minima are degenerate; in C the flights condense at $\lmax$.}
\label{fig:overview}
\end{figure*}

\subsection{One flight}\label{sec:single}

A flight length has no natural unit, so we measure it on a logarithmic scale. Every data set also comes with two bounds: a shortest flight $\lmin$, set by the resolution of the measurement, and a longest one $\lmax$, set by the walker, by the arena or by the duration of the recording. We keep both from the start and work with
\begin{equation}
x=\ln\frac{\ell}{\lmin}\in[0,L],\qquad L=\ln\frac{\lmax}{\lmin},
\label{eq:x}
\end{equation}
so that $L$ counts the $e$-folds spanned by the flights. The measure $dx=d\ell/\ell$ is the only one that a change of units leaves invariant, up to a shift of $x$. Maximizing the entropy $S[p]=-\int_0^Ldx\,p(x)\ln p(x)$ at fixed $\langle x\rangle$ gives
\begin{equation}
p(x)=\frac{e^{\theta x}}{Z(\theta)},\qquad Z(\theta)=\frac{e^{\theta L}-1}{\theta},
\label{eq:single}
\end{equation}
which is a power law $p(\ell)\propto\ell^{-\mu}$ truncated to $[\lmin,\lmax]$, with $\mu=1-\theta$~\cite{Montroll1983}, and not a L\'evy-stable law, which describes the sum of many flights rather than a single one.

For a single flight, $\lmax$ is a convenience. On the half-line $x\ge0$, Eq.~\eqref{eq:single} is a normalized law whenever $\theta<0$, that is, $\mu>1$, and one may let $\lmax\to\infty$. The cutoff does make room for the opposite case: on $[0,L]$ any $\theta$ is allowed, and $\theta>0$ describes a walker whose flights pile up near $\lmax$. Section~\ref{sec:path} shows that, once flights remember each other, this case cannot be avoided, and the cutoff becomes a necessity.

The mean and the variance of $x$ are
\begin{equation}
\begin{aligned}
\langle x\rangle&=\Lambda(\theta)=\frac L2\Big[1+\Lang\Big(\frac{\theta L}2\Big)\Big],\\
\mathrm{Var}\,x&=\Lambda'(\theta)=\frac{L^2}{4}\,\Lang'\Big(\frac{\theta L}2\Big),
\end{aligned}
\label{eq:langevin}
\end{equation}
where $\Lang(y)=\coth y-1/y$ is the Langevin function of paramagnetism [\figref{fig:theory}(a)]. Its appearance has a simple explanation. In the variable $u=2x/L-1\in[-1,1]$ the law~\eqref{eq:single} reads $p(u)\propto e^{bu}$ with $b=\theta L/2$. This is exactly the law of the vertical component $u=\cos\vartheta$ of a classical Heisenberg spin, a unit vector in three dimensions, in a field $b$: the area of a sphere between two heights is proportional to the distance between them (Archimedes' hat-box theorem), so $\cos\vartheta$ is uniform on $[-1,1]$ for a spin without field, and the Boltzmann factor $e^{b\cos\vartheta}$ tilts it as $e^{\theta x}$ tilts the flight. A flight is therefore a classical spin that interacts with nothing but an external field, the elementary unit of Langevin's theory of paramagnetism: the spin points to the south pole when the flight is close to $\lmin$ and to the north pole when it is close to $\lmax$ [\figref{fig:overview}(b)], the field $b$ sets how strongly it is tilted toward one pole or the other, and $\langle u\rangle=\Lang(b)$ is its magnetization. For the planar spins that describe directions of motion in Ref.~\cite{Agliari2024}, the Bessel ratio $I_1/I_0$ plays the same role. The susceptibility $\Lambda'$ is largest at $\theta=0$, where the histogram of $\ln\ell$ is flat and $\Lambda'(0)=L^2/12$; this number will set the critical point of the collective speed in Sec.~\ref{sec:meanfield}.

\subsection{A sequence of flights}\label{sec:path}

A walker produces flights $\mathbf x=(x_1,\dots,x_T)$ one after the other [\figref{fig:overview}(a)]. Following Ref.~\cite{Agliari2024}, in the spirit of maximum caliber~\cite{Presse2013}, we maximize the entropy of the whole sequence, $S[P]=-\int\mathcal Dx\,P[\mathbf x]\ln P[\mathbf x]$ with $\mathcal Dx=\prod_tdx_t$, at fixed mean of $x$ and fixed lagged correlations
\begin{equation}
C(\tau)=\frac1{T-K}\sum_{t=K+1}^T\langle x_tx_{t-\tau}\rangle,\qquad\tau=1,\dots,K .
\end{equation}
The lagged correlations are the constraints that carry the memory. Fixing only the mean returns the law~\eqref{eq:single} for every flight, independently of the others, so that a walker described by it has no memory by construction. Each lagged correlation that is also fixed lets a flight depend on the one $\tau$ steps earlier, and on nothing else that the data have not demanded; whether the data do demand it, and up to which lag $K$, is then a question that the data answer (Sec.~\ref{sec:estimation}). In this way MaxEnt turns the question ``does the walker remember?'' into the estimate of a few multipliers. Introducing the Lagrange multipliers $h$ for the mean and $J_\tau$ for the correlation at lag $\tau$, the law of maximum entropy is
\begin{equation}
P[\mathbf x]=\frac1{Z_T}\exp\Big[\sum_tx_t\Big(h+\sum_{\tau=1}^KJ_\tau\,x_{t-\tau}\Big)\Big],
\label{eq:joint}
\end{equation}
the Boltzmann weight of a one-dimensional chain of continuous spins in time, in a field $h$ and with couplings $J_\tau$ of range $K$ [\figref{fig:overview}(b)]. \rev{As in $C(\tau)$, the sum runs over $t=K+1,\dots,T$, the first $K$ flights being held fixed.} For $K=0$ the flights are independent and identically distributed: this is the renewal L\'evy walk. A positive coupling means that long flights tend to follow long flights, a negative one that they alternate with short ones. With Gaussian instead of scale-free constraints the same construction gives Burg's maximum-entropy autoregression~\cite{Burg1975,CoverThomas}.

This is where the upper cutoff earns its place. Seen from flight $t$, the past acts as an effective field $h+\sum_\tau J_\tau x_{t-\tau}$. With $J_\tau>0$ this field grows with the lengths of the previous flights, and a run of flights that are long enough makes it positive, so that the next flight favors long lengths. On the half-line there is no normalizable law $e^{\theta x}$ with $\theta>0$, and the weight~\eqref{eq:joint} itself cannot be normalized. On $[0,L]$ every history yields a proper law, and every value of the parameters is allowed. What the cutoff costs is a scale: the largest change of field that one flight can produce on another is $J_\tau L$, and it is this product, not $J_\tau$ alone, that measures the strength of the memory (Sec.~\ref{sec:condensation}).

When several walkers are observed together, we give each walker $i$ its own field $h_i$. The fields absorb the differences between individuals, a quenched random field in the language of disordered systems, while the couplings, which describe the memory, are shared.

\subsection{The causal chain}\label{sec:causal}

The normalization $Z_T$ couples all the flights of Eq.~\eqref{eq:joint}, which makes its likelihood costly to evaluate. Agliari \emph{et al.} removed the difficulty with a mean-field step~\cite{Agliari2024}: in the weight of each spin they replaced the field generated by the past with the value it takes on the observed data, and normalized spin by spin, which makes the measure factorize in time. For one walker this step has an exact meaning. Each factor becomes the law of the next flight given the observed previous ones,
\begin{equation}
p(x_t\,|\,x_{t-1},\dots,x_{t-K})=\frac{e^{\theta_tx_t}}{Z(\theta_t)},\qquad
\theta_t=h+\sum_{\tau=1}^KJ_\tau\,x_{t-\tau},
\label{eq:causal}
\end{equation}
and their product is the probability of the observed sequence under a Markov chain of order $K$. This chain is a MaxEnt object in its own right. It maximizes, under the same constraints, the causal entropy
\begin{equation}
S_{\rm c}[P]=-\sum_t\int\mathcal Dx\,P[\mathbf x]\ln p(x_t\,|\,x_{t-1},\dots,x_{t-K}),
\label{eq:causalentropy}
\end{equation}
the sum of the entropies of each flight given the $K$ previous ones~\cite{Ziebart2013}, and it is the model we use. At each turn the walker draws a truncated power law whose exponent, $\mu_t=1-\theta_t$, is set by its recent flights.

The log-likelihood
\begin{equation}
l(h,\bm J)=\sum_t\big[\theta_tx_t-\ln Z(\theta_t)\big]
\label{eq:loglik}
\end{equation}
\rev{summed over the same flights as Eq.~\eqref{eq:joint}, or over those with at least $K_{\max}$ predecessors when depths up to $K_{\max}$ are compared (Appendix~\ref{app:estimation}),} is an elementary function of the parameters $(h,\bm J)$, and it is concave in them, because $\ln Z$ is a convex function of $\theta$ and $\theta_t$ is linear in $(h,\bm J)$. Its maximum, when it exists, is unique. The stationarity conditions,
\begin{equation}
\sum_t\big[x_t-\Lambda(\theta_t)\big]=0,\qquad\sum_t\big[x_t-\Lambda(\theta_t)\big]\,x_{t-\tau}=0,
\label{eq:score}
\end{equation}
are the dynamic version of moment matching: the difference between each flight and its predicted mean must vanish on average and be uncorrelated with the past. The Fisher information,
\begin{equation}
\mathsf I=\sum_t\Lambda'(\theta_t)\,\bm z_t\bm z_t^{\mathsf T},\qquad \bm z_t=(1,x_{t-1},\dots,x_{t-K}),
\label{eq:fisher}
\end{equation}
gives the error bars. No Gaussian surrogate of the likelihood is needed: Eq.~\eqref{eq:loglik} is maximized as it is.

The path measure~\eqref{eq:joint} and the causal chain~\eqref{eq:causal} are different families with the same sufficient statistics. They agree to first order in $J$, after a shift of the field, and fitted to the same flights they return nearly the same couplings (Appendix~\ref{app:single}). In both, the selected depth is a Markov order~\cite{Ge2012}: $K^*=0$ is renewal, $K^*=1$ a Markov chain of flights, $K^*>1$ a longer memory. We call
\begin{equation}
\mub=1-h
\end{equation}
the bare exponent. It is the exponent with which the walker draws its next flight after a run of the shortest possible ones ($x_{t-\tau}=0$ for every $\tau$), when the memory adds nothing to the field. With memory it differs from the exponent of the histogram of all flights (Sec.~\ref{sec:dressing}).

\begin{figure*}[t]
\centering
\includegraphics[width=\textwidth]{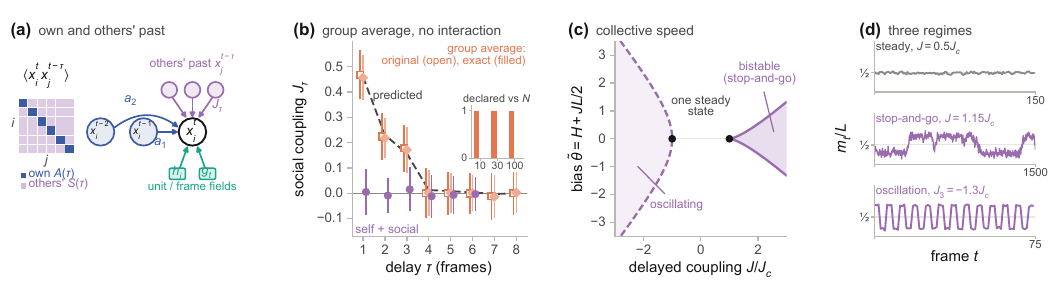}
\caption{Many walkers. (a)~The lagged correlation between units has a diagonal part, each unit with its own past ($A$), and an off-diagonal part, each unit with the others' past ($S$); the group average mixes the two, Eq.~\eqref{eq:Rsplit}. The self\,+\,social model~\eqref{eq:selfsocial} gives each part its own couplings, $a_\tau$ and $J_\tau$, and adds a field for each unit, $h_i$, and for each frame, $g_t$. (b)~Synthetic units with own memory of depth 3 and no interaction ($N=30$, 40 replicates, mean $\pm$ s.d.): couplings returned by the group-average analysis in its original implementation (open squares) and with its exact likelihood (filled diamonds), against the prediction of Eq.~\eqref{eq:aggregation} (dashed); circles, social couplings of the self\,+\,social model. Inset: fraction of data sets in which a social coupling is declared, for $N=10$, 30 and 100. (c)~Mean-field phase diagram of the collective speed with one delayed coupling $J$, in units of $J_c=12/L^2$, and bias $\tilde\theta$: two stable states (shaded, solid boundary) for $J>J_c$, oscillations (dashed boundary) for $J<-J_c$; dots, the critical points. (d)~Group average on the symmetric line $\tilde\theta=0$: one steady state ($J=0.5J_c$, $N=200$), collective stop-and-go ($J=1.15J_c$, $N=40$), and oscillations of period 6 for a negative coupling at delay 3 ($N=200$). $L=\ln100$ throughout.}
\label{fig:many}
\end{figure*}

\subsection{Many walkers on a common clock}\label{sec:many}

In a group, the flights of different individuals start and end at different times, so ``the previous flight of the others'' is not defined. A group needs a common clock. We use frames $t$ of fixed duration and, for unit $i$, the log-displacement per frame, or log-speed, $x_i^t=\ln(\ell_i^t/\lmin)\in[0,L]$, with the same single-site law~\eqref{eq:single}.

Frames and flights are two different observables, and they answer different questions. On a frame clock a single flight spans many frames, so a walker's memory from one frame to the next mostly measures how long it keeps moving fast or slow; a renewal L\'evy walk observed in frames is strongly correlated, simply because its flights last. Whether a walker is a renewal process can only be asked of its flights, the events at which it chooses anew, and that is what we do for single walkers. To ask whether a walker responds to the others, instead, we must know what each of the others was doing $\tau$ seconds before the walker's step. Flight indices do not provide this, since the $t$-th flight of one walker and the $t$-th flight of another happen at different times; a clock shared by all walkers does, because it gives every walker a state at every instant. For groups we therefore use frames. The model is the same; only the index of time changes.

Transcribing Eq.~(A16) of Ref.~\cite{Agliari2024} from directions to log-speeds gives the path measure
\begin{equation}
P(\mathbf x)=\frac1Z\exp\Big[\sum_{t,i}x_i^t\Big(H+\frac1N\sum_{\tau=1}^{\tau_M}J_\tau\sum_jx_j^{t-\tau}\Big)\Big],
\label{eq:pre}
\end{equation}
which fixes the mean of $x$ and the lagged autocorrelation $R(\tau)=\langle m_tm_{t-\tau}\rangle$ of the group average $m_t=N^{-1}\sum_ix_i^t$. This autocorrelation contains two different things,
\begin{equation}
R(\tau)=\frac1NA(\tau)+\Big(1-\frac1N\Big)S(\tau),
\label{eq:Rsplit}
\end{equation}
where $A(\tau)$ is the mean lagged correlation of each unit with its own past and $S(\tau)$ that with the past of the others [\figref{fig:many}(a)]. In Eq.~\eqref{eq:pre} one multiplier $J_\tau$ accounts for both. Giving each its own multiplier, and each unit and each frame its own field, yields the self\,+\,social model,
\begin{equation}
\theta_i^t=h_i+g_t+\sum_{\tau=1}^{K_s}a_\tau\,x_i^{t-\tau}+\sum_{\tau=1}^{K_c}J_\tau\,s_i^{t-\tau},
\label{eq:selfsocial}
\end{equation}
with $p(x_i^t\,|\,\text{past})=e^{\theta_i^tx_i^t}/Z(\theta_i^t)$ and units independent of each other given the past of all. The self couplings $a_\tau$ capture individual persistence, playing the same role as the memory couplings in the single-walker model (Sec.~\ref{sec:causal}). The social couplings $J_\tau$ act on $s_i^{t-\tau}$, the mean log-speed at frame $t-\tau$ of all other units (global coupling) or of the $k$ units nearest to $i$ (local coupling). The time fields $g_t$ absorb whatever affects all units at once. The model of Ref.~\cite{Agliari2024} is the special case $h_i=H$, $g_t=0$, $a_\tau=J_\tau/N$, $s=m$.

In Eq.~\eqref{eq:selfsocial} a unit's own past and the past of its neighbors enter in exactly the same way, as a field on its next step weighted by a coupling. Formally, the unit's own previous step is one more neighbor, the one that is always at distance zero: a memory is an interaction with one's own past, and an interaction is a memory of somebody else's. The two differ in what can produce them. Own memory can come from inertia, physiology or the persistence of a behavioral state; a coupling to the others' past requires that the unit perceives them. A drive common to all units, such as a change of temperature, also correlates them, but it is neither memory nor interaction, and the time fields $g_t$ are there to absorb it (Sec.~\ref{sec:groupavg}). Own memory and social coupling are kept in separate terms because the data offer them separately, in the diagonal and off-diagonal blocks of the lagged correlations of Fig.~\ref{fig:many}(a). For the same reason the single-walker model is not the trivial case $N=1$ of the collective one: it is the diagonal term that must be modeled before the off-diagonal one can be read.

{Directions of motion are modeled in the same way, with planar instead of Heisenberg spins. The heading of unit $i$, a unit vector $\bm n_i^t$, is drawn from the von Mises law $p(\bm n_i^t\,|\,\text{past})=e^{\bm\theta_i^t\cdot\bm n_i^t}/[2\pi I_0(|\bm\theta_i^t|)]$, the planar spin of Ref.~\cite{Agliari2024}, with the field $\bm\theta_i^t=\mathbf h_i+\mathbf g_t+\sum_\tau a_\tau\bm n_i^{t-\tau}+\sum_\tau J_\tau\bm s_i^{t-\tau}$ built as in Eq.~\eqref{eq:selfsocial} from fields in the plane and from the mean heading $\bm s_i$ of the neighbors (Appendix~\ref{app:collective}). The modulus $|\bm\theta_i^t|$ is the concentration of the law: for small deviations, the inverse square of the angular spread of the heading about the direction of $\bm\theta_i^t$.}

\section{Statistical mechanics of the memory chain}\label{sec:statmech}

\subsection{The stationary law}\label{sec:stationary}

With one lag, $\theta(x)=h+Jx$, the causal chain~\eqref{eq:causal} satisfies detailed balance with respect to
\begin{equation}
\pi(x)\propto e^{hx}\,Z(h+Jx),\qquad \pi(x)\,p(y\,|\,x)\propto e^{h(x+y)+Jxy},
\label{eq:stationary}
\end{equation}
as one checks by inserting Eq.~\eqref{eq:single} (Appendix~\ref{app:single}). The joint law of two consecutive flights is a two-site Boltzmann weight of Ising type, with continuous spins on $[0,L]$, and every stationary property of the walk follows from it.

The stationary law has a simple reading. In a reversible chain, the weight of a flight $x$ is the total weight of the pairs it forms with its successor, $\int_0^Ldy\,e^{h(x+y)+Jxy}=e^{hx}Z(h+Jx)$. The first factor is the bare power law. The second counts the successors: for $J>0$ a long flight forms favorable pairs with long successors, so the factor $Z(h+Jx)$ grows with $x$, roughly as $e^{JxL}$ once the field $h+Jx$ is positive, and it lifts the tail of the distribution. Figure~\ref{fig:overview}(d) shows the effect for three chains, marked A, B and C in the phase diagram of \figref{fig:overview}(c). They share the bare exponent, $\mub=2$, and differ only in the coupling, $J_1L=0.5$, 1.5 and 2.5, and each gives one curve in \figref{fig:overview}(d). For chain A, with the weakest coupling, the stationary law is barely distinguishable from the renewal one; for chain B the tail bends upward before the cutoff; for chain C most flights sit at $\lmax$. The three chains lie in three different regions of the phase diagram, and the next section explains what separates them.

\subsection{Duality and condensation}\label{sec:condensation}

Section~\ref{sec:path} showed that, once flights remember each other, the upper cutoff cannot be removed: a positive coupling makes long flights favor long successors, and only $\lmax$ keeps them finite. What, then, does a positive memory do to a walk? It may leave a power-law histogram, only heavier-tailed, as for chain A; or runs of long flights may take over and pile the flights up at $\lmax$, as for chain C. The question matters for data. A walker on the far side of this change would no longer look like a L\'evy walker at all, and the location of the change is the scale against which the coupling of a real walker is strong or weak. We answer it with three standard tools of statistical mechanics. A symmetry of the pair weight, a duality, locates the change exactly without solving the chain, as the Kramers--Wannier duality locates the critical temperature of the two-dimensional Ising model~\cite{KramersWannier1941}. For wide ranges a large-deviation estimate of the stationary law yields a free energy, whose minima show that the change is a first-order transition. Finite ranges, those of the data, are then treated exactly with Eq.~\eqref{eq:stationary}.

Exchange long and short flights, $\ell\to\lmin\lmax/\ell$, that is, $x\to L-x$. A flight close to $\lmin$ becomes one close to $\lmax$ and vice versa, while the geometric mean $\sqrt{\lmin\lmax}$, the midpoint $x=L/2$ of the logarithmic range, stays where it is. For a renewal walker this maps the exponent $\mu$ onto $2-\mu$, a law that favors short flights onto one that favors long ones. With memory, the pair weight $e^{h(x+y)+Jxy}$ keeps its form: the coupling is unchanged and the field becomes $-h-JL$, up to a constant. A chain at $(h,J)$ and one at $(-h-JL,J)$ are thus the same chain read upside down: whatever the first does near $\lmin$, the second does near $\lmax$. We call two such chains dual to each other, and a chain that is its own dual self-dual. The self-dual chains are those with $h=-JL/2$. Their stationary law is symmetric about the midpoint, so exactly half of their flights are longer than $\sqrt{\lmin\lmax}$, for every range. With $\mub=1-h$, the self-dual line is the coupling
\begin{equation}
\Jcond=\frac{2(\mub-1)}{L}=2\Jup,\qquad \Jup=\frac{\mub-1}{L}.
\label{eq:thresholds}
\end{equation}

Both thresholds have a direct meaning. Take $\mub>1$, a rule that by itself favors short flights. After a flight at the cutoff, $x=L$, the field is $h+JL=JL-(\mub-1)$, and it is positive, so that the next flight is more likely long than short, when $J>\Jup$. Above this coupling a run of long flights can sustain itself. Whether the walker spends most of its time in such runs is decided by the duality: below $\Jcond$ it lives among short flights, above it among long ones. The natural order parameter is therefore the fraction of flights longer than the fixed point of the duality,
\begin{equation}
\Pi=\int_{L/2}^{L}dx\,\pi(x)=\mathrm{Prob}\big(\ell>\sqrt{\lmin\lmax}\big),
\label{eq:order}
\end{equation}
computed with the exact stationary law~\eqref{eq:stationary}. The duality maps $\Pi$ onto $1-\Pi$, so that $\Pi=1/2$ on the self-dual line, and $\Pi$ tends to 0 or to 1 on either side of it. The ratio
\begin{equation}
\Phi=\frac J\Jup=\frac{JL}{\mub-1}
\end{equation}
is the natural control parameter, with $\Phi=1$ at the first threshold and $\Phi=2$ at the second.

For wide ranges the change becomes a sharp transition. From Eq.~\eqref{eq:single}, $Z(\theta)\simeq e^{\theta L}/\theta$ for $\theta L\gg1$, $Z(\theta)\simeq1/|\theta|$ for $\theta L\ll-1$ and $Z(0)=L$, so that
\begin{equation}
\ln Z(\theta)=L\max(\theta,0)+O(\ln L):
\end{equation}
the normalization of one flight is dominated by flights near $\lmax$ when its field is positive and by flights near $\lmin$ when it is negative (Appendix~\ref{app:largeL}). Inserting this into $\pi(x)\propto e^{hx}Z(h+Jx)$, at fixed $\alpha=\mub-1>0$ and $\beta=JL$, shows that the stationary law obeys a large-deviation principle in $L$,
\begin{equation}
\pi(x)\asymp e^{-L\,F(x/L)},\qquad F(u)=\alpha u-\max(\beta u-\alpha,0),
\label{eq:freeenergy}
\end{equation}
with $u=x/L\in[0,1]$ [\figref{fig:overview}(e)]. The logarithmic range $L$ plays the role of system size and $F$ that of a free energy per unit size. Because both factors of $\pi$ are exponentials in $x$, $F$ is piecewise linear and its minima can only sit at the ends, $u=0$ (flights near $\lmin$) and $u=1$ (flights near $\lmax$). For $\beta<2\alpha$ the global minimum is $u=0$ and $\Pi\to0$: the walker is in a Pareto phase, with a power-law histogram. For $\beta>2\alpha$ the global minimum is $u=1$ and $\Pi\to1$: the flights condense at the cutoff. The transition, on the self-dual line, is first order, with $\Pi$ jumping from 0 to 1. Between the two thresholds, $\alpha<\beta<2\alpha$, the end $u=1$ is a local minimum: a run of flights near $\lmax$ is metastable, and it lasts for a number of flights that grows exponentially with $L$ (Appendix~\ref{app:largeL}); $J=\Jup$ is the spinodal. At finite $L$ the jump is smoothed over a width of order $1/L$, and the curves $\Pi(\Phi)$ for different ranges cross at the self-dual point $\Phi=2$, $\Pi=1/2$ [\figref{fig:theory}(b)], as observables do at first-order transitions in finite systems~\cite{BinderLandau1984}.

Both thresholds vanish as $\lmax/\lmin\to\infty$ and this reflects how the coupling scales. The coupling acts through $JL$, the largest change of field that one flight can impose on the next, and it is $JL$ that must be held fixed as the range grows, much as the couplings of a mean-field magnet must scale with the inverse number of spins. At fixed $J$ a wider range makes the same coupling stronger, and on the half-line every positive coupling lies beyond condensation, which is another way of saying that the chain does not exist there. At fixed $\Phi$ the limit $L\to\infty$ is well defined, and it is the limit in which the transition becomes sharp. The cutoff is therefore a physical ingredient of a L\'evy walk with memory, not a regularization to be removed at the end. Real data span one to two and a half decades, where the transition is a broad crossover, but its location still organizes the phase diagram in which the walkers of Sec.~\ref{sec:data} sit [\figref{fig:overview}(c)].

\rev{Nothing in the duality requires a single lag or detailed balance. Since $Z(-\theta)=e^{-\theta L}Z(\theta)$, exchanging long and short flights, the next one and all the previous ones, maps the causal kernel~\eqref{eq:causal} of any depth onto itself, with the couplings unchanged and the field $h\to-h-J_{\rm tot}L$, where $J_{\rm tot}=\sum_\tau J_\tau$. For $K>1$ the chain is not reversible, and its stationary law, which is unique, has no closed form. Yet the self-dual chains, $J_{\rm tot}=2(\mub-1)/L$, still have a stationary law symmetric about $L/2$, and therefore $\Pi=1/2$; and with non-negative couplings a run of flights at the cutoff sustains itself when $J_{\rm tot}>(\mub-1)/L$, with $\mub=1-h$ the exponent after a run of the shortest flights, as before. The two lines of the phase diagram of \figref{fig:overview}(c) thus apply at any depth, with $J_{\rm tot}$ in place of $J_1$ and $\Phi=J_{\rm tot}L/(\mub-1)$, and only its shading, computed from the exact law~\eqref{eq:stationary}, is specific to $K=1$; the same total coupling controls the collective speed (Sec.~\ref{sec:meanfield}). A fit with $K=1$ to a walker with a deeper, positive memory returns a lag-1 coupling that absorbs part of the deeper couplings, larger than $J_1$ and smaller than $J_{\rm tot}$, so that its coordinates place the walker short of where it actually sits (Sec.~\ref{sec:termites}).}

\rev{The stationary law counts flights. A walker that moves at constant speed spends in each flight a time proportional to its length, so the fraction of time spent in flights of log length $x$ is the length-biased law $\pi_{\rm time}(x)\propto e^x\pi(x)\propto e^{(h+1)x}Z(h+Jx)$, familiar from renewal theory~\cite{Zaburdaev2015}, with free energy $F(u)-u$. The bare factor is that of a walker with exponent $\mub-1$; the memory factor is unchanged. For wide ranges, a walker with $\mub<2$ spends most of its time in flights near $\lmax$ even without memory, the ballistic regime of the renewal L\'evy walk. For $\mub\ge2$ time condenses at $JL=2(\mub-1)-1$, that is, at $\Phi=2-1/(\mub-1)$, between the spinodal and the condensation of the flights: measured in time rather than in flights, runs of long flights take over earlier.}

\subsection{Memory dresses the exponent}\label{sec:dressing}

A histogram does not know in which order the flights came. An i.i.d.\ truncated-Pareto fit to stationary flights matches their mean log length and returns $\hat\mu=1-\Lambda^{-1}(\langle x\rangle_\pi)$. For $J>0$ this exponent is smaller than $\mub$ [\figref{fig:theory}(c)]: memory packs long flights into runs, the histogram is heavier-tailed than the rule that generates it, and a walker with memory looks more L\'evy than its one-step rule. On the self-dual line $\langle x\rangle_\pi=L/2$ and $\hat\mu=1$ exactly, for every range. \rev{Beyond it $\hat\mu<1$, and deep in the condensed phase $\hat\mu<0$ [\figref{fig:theory}(c)]. On $[0,L]$, an effective parameter $\hat\mu=1-\hat\theta<1$ describes a histogram that rises toward $\lmax$, $\hat\theta>0$, a law that the half-line cannot host and that the condensed phase requires.} Put differently, the local exponent $\mu_t=1-\theta_t$ fluctuates with the recent past, a superstatistics~\cite{BeckCohen2003} generated by the dynamics itself. The exponent of a histogram is thus not a property of the search rule alone; it also depends on the memory and on the range over which it is measured. In real walkers the difference exceeds half a unit (Sec.~\ref{sec:data}).

\subsection{A critical point for the collective speed}\label{sec:meanfield}

Groups raise a question that single walkers do not: can a delayed coupling between walkers order the group as a whole? Take $N$ units with a global delayed coupling and no own memory, that is, Eq.~\eqref{eq:selfsocial} with $a_\tau=0$, $s_i$ the mean of all other units, a common field $h_i=H$ and no time fields. The group average $m_t$ plays the role of a magnetization. It is the collective log-speed, close to $0$ when the whole group crawls and close to $L$ when it runs. Two parameters control it. One is the total coupling $J_{\rm tot}=\sum_\tau J_\tau$, the strength with which the past speed of the group feeds back on its present. The other is a bias. Exchanging slow and fast, $m\to L-m$, maps the model onto itself with $H\to-H-J_{\rm tot}L$, the duality of Sec.~\ref{sec:condensation} at the level of the group; its symmetric point is $H=-J_{\rm tot}L/2$, and
\begin{equation}
\tilde\theta=H+\frac{J_{\rm tot}L}2
\end{equation}
measures the distance from it, positive when the group is biased toward running.

For $N\to\infty$ the units are independent given the past, and $m_t$ follows the deterministic delayed map
\begin{equation}
m_t=\Lambda\Big(H+\sum_\tau J_\tau m_{t-\tau}\Big),
\label{eq:map}
\end{equation}
with $\Lambda$ the Langevin response~\eqref{eq:langevin}. A steady state $m^*$, with field $\theta^*$, loses stability when a root $z$ of $1=\Lambda'(\theta^*)\sum_\tau J_\tau z^{-\tau}$ crosses the unit circle (Appendix~\ref{app:collective}). For positive couplings the first root to cross is $z=1$, a fold at which a second stable state appears. Since the susceptibility of one flight never exceeds $L^2/12$ (Sec.~\ref{sec:single}), this requires
\begin{equation}
J_{\rm tot}>J_c=\frac{12}{L^2},
\label{eq:Jc}
\end{equation}
with a cusp at $J_{\rm tot}=J_c$, $\tilde\theta=0$ [\figref{fig:many}(c)]. Above $J_c$ the group switches at random between a slow and a fast state, a collective stop-and-go whose residence times grow with $N$ [\figref{fig:many}(d)]. A negative coupling at delay $d$ makes the group oscillate instead: the map splits into $d$ independent sub-chains, each of which undergoes a period doubling at $|J|=J_c$, and any lag-1 inertia puts them in phase, so that the period is $2d$. Near both thresholds the fluctuations of the group average diverge, $N\,\mathrm{Var}(m)\to\Lambda'/[1-(\Lambda'J)^2]$ for one lag.

Own memory adds to the social feedback. With a lag-1 self coupling $a\ge0$, each unit at a steady state is the chain of Eq.~\eqref{eq:stationary} in the field $H+J_{\rm tot}m^*$, and the steady state loses stability when $J_{\rm tot}\chi=1$, where $\chi=\sum_{s\ge0}R(s)$ is the integrated response of one unit to a pulse of field. What enters is the response, not the integrated autocorrelation, which is 50\% larger at $a=0.2$. For $L=\ln100$ and $a=0.2$ the threshold drops to $J_c(a)=0.71J_c$, against $0.65J_c$ from the naive estimate $\Lambda'(0)(a+J)=1$, and the theory reproduces the simulated fluctuations within 1\% away from the threshold and within 4\% at $J=0.6J_c$ [Appendix~\ref{app:collective}, \figref{fig:collective}(b)].

The threshold $J_c=12/L^2$ exists because the support of each speed is compact, and it vanishes as the range grows. It is a scale-free and delayed counterpart of the speed criticality reported for starling flocks~\cite{Bialek2014}, where the interactions are instantaneous.

\section{From trajectories to couplings}\label{sec:inference}

\subsection{Fitting and choosing the depth}\label{sec:estimation}

For $N$ walkers, $i=1,\dots,N$, with flights $x_{i,t}$, the field of Eq.~\eqref{eq:causal} becomes $\theta_{i,t}=h_i+\sum_\tau J_\tau x_{i,t-\tau}$, with one field per walker and shared couplings. The log-likelihood stays concave, and Newton's method finds its maximum in a few iterations (Appendix~\ref{app:estimation}). The cost is negligible. For one larval condition (6 to 59 larvae, 349 to 4240 flights) the whole scan of depths $K=0,\dots,6$, with one field per larva, takes 0.1--0.6\,s on a single core of a desktop computer; pooling all 241 larvae, with 241 fields and up to six couplings for 20\,948 flights, takes 5\,s. The same machinery accommodates fields that change every $B$ flights, which absorb slow drifts of the exponent and leave to the couplings only the short-range memory, and known covariates, such as the distance from a wall. Short tracks bias the couplings of a fit with one field per walker by an amount of order $1/T_i$~\cite{Nickell1981}, which we measure on surrogates in which the flights of each walker are permuted.

The depth is chosen by comparing models with $K=0,\dots,K_{\max}$, fitted on the same flights, through an information criterion, the Akaike information criterion, $\mathrm{AIC}=-2l+2k$, or the Bayesian information criterion, $\mathrm{BIC}=-2l+k\ln n_{\rm o}$, with $k$ parameters and $n_{\rm o}$ predicted flights~\cite{Akaike1974,Schwarz1978}. The choice is not innocent. On renewal data AIC selects $K^*>0$ with asymptotic probability
\begin{equation}
P_{\rm AIC}=1-\exp\Big[-\sum_{k=1}^{K_{\max}}\frac1k\,P(\chi^2_k>2k)\Big],
\label{eq:woodroofe}
\end{equation}
equal to 0.248, 0.267, 0.284 and 0.288 for $K_{\max}=4$, 6, 12 and $\infty$, however long the data. This classical consequence of Spitzer's combinatorial lemma~\cite{Shibata1976,Woodroofe1982} makes AIC inconsistent for the order of a chain, and cross-validation with a fixed training fraction, asymptotically equivalent to it~\cite{Stone1977,Shao1993}, inherits the same bias. BIC is consistent~\cite{Katz1981,CsiszarShields2000}. Heavy tails play no role here, since in the log variable every sufficient statistic is bounded. We select the depth with BIC and calibrate every claim against surrogate data.

On walkers generated by the model itself with random parameters, the procedure does what it should (Appendix~\ref{app:selection}, \figref{fig:validation}). Planted and inferred couplings agree, and their differences, in units of the Fisher errors~\eqref{eq:fisher}, have mean $-0.01$ and standard deviation 0.97. BIC recovers the planted depth in 78\% of the walkers and never selects a deeper chain, whereas AIC does so in 24\%. On renewal walkers the false-positive rate of AIC stays at the level of Eq.~\eqref{eq:woodroofe} from 250 to 16\,000 flights, and that of BIC vanishes as the data grow.

\subsection{What else looks like memory}\label{sec:masquerade}

A real trajectory can mimic memory in ways that a planted walker cannot, and three of them matter here (Appendix~\ref{app:pitfalls}, \figref{fig:pitfalls}). Walkers differ: two flights of the same fast walker are both long, and a model with one field for all reads this as memory, at the largest depth allowed. One field per walker removes the effect, which is why we never pool walkers without their own fields. Flights are not observed but cut out of noisy tracks, and noise occasionally breaks a long flight into two long pieces, which manufactures a positive correlation; merging consecutive pieces that are nearly collinear undoes it, and a pipeline null, which rebuilds each track from its own flights in random order and analyzes it again, measures whatever memory the pipeline itself creates. Finally, the variable matters: the linear statistic $\ell_t\ell_{t-1}$ is dominated by a handful of extreme flights and misses a memory that the log statistic $x_tx_{t-1}$ detects. Different mechanisms of dependence also leave different signatures in how the selected depth grows with the length of the data, which helps to tell a finite memory from a slowly decaying one.

\subsection{Reading interactions from a group}\label{sec:groupavg}

For groups, three results, proved in Appendix~\ref{app:collective} and tested on synthetic collectives in Table~\ref{tab:collective}, organize the inference.

\emph{The mean-field measure is the exact likelihood of a causal model.} The factorized measure of Ref.~\cite{Agliari2024}, evaluated on the data, is a product of normalized laws, one for each unit and frame, with the field $\theta_t=H+\sum_\tau J_\tau m_{t-\tau}$ computed from the observed past. It is therefore exactly the likelihood of the causal chain with this field. This chain is not the path measure~\eqref{eq:pre}, just as for single walkers the causal chain is not the path measure (Sec.~\ref{sec:causal}); but it is a well-defined model whose likelihood can be maximized exactly, with no further approximation. The Gaussian autoregressive moment approximation of Ref.~\cite{Agliari2024} overestimates planted couplings by 24\% in groups of five units, whereas the exact likelihood estimator remains unbiased across group sizes.

\emph{The group average contains own memory.} Let the units be independent, each with its own memory and autocovariance $C_i(\tau)$. A model of the group average then acquires delayed couplings, because the autocorrelation of the average contains the autocorrelation of each unit, Eq.~\eqref{eq:Rsplit}. Linearizing the inversion of Ref.~\cite{Agliari2024}, the fitted couplings converge to
\begin{equation}
\begin{gathered}
J^\star=\frac{\phi}{\Lambda'(\bar\theta)},\qquad\phi=\bar{\mathsf R}^{-1}\bar{\bm c},\\
\bar{\mathsf R}_{\tau\tau'}=\bar C(|\tau-\tau'|),\qquad \bar c_\tau=\bar C(\tau),
\end{gathered}
\label{eq:aggregation}
\end{equation}
the Yule--Walker coefficients of the mean individual autocovariance $\bar C=N^{-1}\sum_iC_i$, divided by the susceptibility of one unit. The factors of $N$ cancel, so these couplings do not shrink with the size of the group. In simulations without interactions the group-average analysis declares a delayed coupling in every data set, for $N=10$, 30 and 100 alike, with coefficients that follow Eq.~\eqref{eq:aggregation} lag by lag [\figref{fig:many}(b)]: $(0.468,0.225,0.177)$ against the predicted $(0.459,0.230,0.156)$ at $N=30$. These couplings are not an error of the analysis. They are the memories of the units, measured through their average, and a unit's own past is as legitimate a source of delayed correlation as the past of its neighbors; but the two are memories of a different nature, and the average cannot separate them. In the validation of Ref.~\cite{Agliari2024} on the Vicsek model, a peak at the first delay was correctly attributed to the inertia of each particle; Eq.~\eqref{eq:aggregation} shows that an inertia lasting longer than one frame produces peaks at longer delays as well. The self\,+\,social model assigns own memory to the couplings $a_\tau$: with regressors that leave out the unit itself, its social score vanishes exactly when units do not interact, and it declares a coupling in 1 of 120 interaction-free data sets.

\emph{Common drives look like interactions.} A field shared by all units, such as temperature, illumination or a pulse of chemokine, correlates units that never interact. Time fields $g_t$ absorb it, but they also absorb a global coupling: once each frame has its own field, the mean of the other units at an earlier frame is a function of that frame plus a multiple of the unit's own past, and global coupling and common drive become indistinguishable. A local coupling, which differs from unit to unit within a frame, survives. Local coupling with time fields is the only reading that declares no coupling under a drive alone and recovers a planted coupling without bias (Appendix~\ref{app:collective}, \figref{fig:collective}).

One last control is specific to real groups. In a time-shift surrogate each unit's track is displaced in time by an independent random offset. The surrogate keeps every unit's own memory, its heterogeneity and its marginal law, and it destroys every synchrony between units, both interactions and common drives. A coupling that survives in the surrogates is not an interaction.

\section{Walkers and groups in the data}\label{sec:data}

\begin{figure*}[t]
\centering
\includegraphics[width=\textwidth]{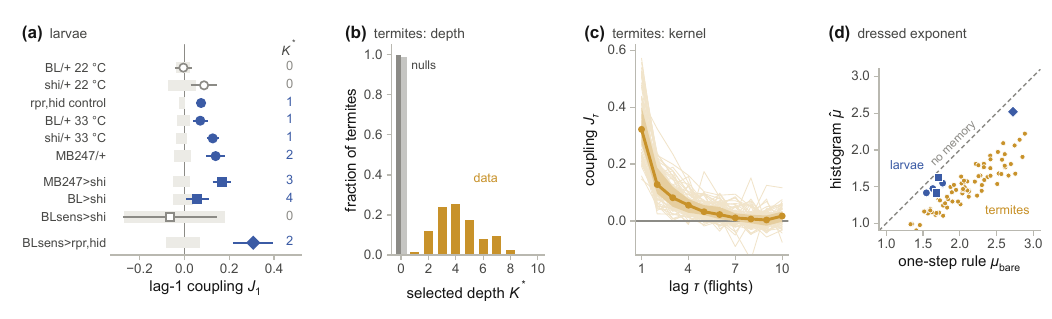}
\caption{Single walkers. (a)~\emph{Drosophila} larvae: lag-1 coupling with one field per larva ($\pm2$ s.e.) and, on the right, the depth $K^*$ selected by BIC; filled symbols, $K^*>0$; grey bands, $\pm2$ s.d.\ over 30 within-larva permutations of the flights. Circles, controls; squares, neurons silenced with \emph{shibire}$^{ts}$; diamond, brain lobes and sensory neurons ablated (genotypes in Appendix~\ref{app:data}). (b)~Depth selected in each of 75 termites (ochre), and by the within-track permutation and pipeline nulls (grey), which select $K^*=0$. (c)~Memory kernel of the termites, from fits with $K=10$: single termites (thin lines), median and interquartile range. (d)~Exponent of the histogram of all flights, $\hat\mu$, against the exponent of the one-step rule, $\mub$, for the larval conditions with memory (medians over larvae, symbols as in (a)) and for each termite; dashed, the diagonal of renewal walkers.}
\label{fig:walkers}
\end{figure*}

We now apply the estimator of Sec.~\ref{sec:inference} to five open data sets (Table~\ref{tab:datasets}), in two steps. Single walkers come first: \emph{Drosophila} larvae and termites, analyzed on the clock of their flights, where we ask whether each walker is a renewal process, how deep its memory is, and where it sits in the phase diagram of Sec.~\ref{sec:statmech}. Groups follow: wild baboons and homing pigeons, analyzed on a common clock of frames, where we ask whether a unit responds to the past of its neighbors once its own memory and the drives common to the group are accounted for. Zebrafish T cells close the section as a negative control for both questions.

All single walkers go through the same pipeline (Appendix~\ref{app:data}): flights are cut out of each track with a corridor of half-width $\varepsilon=0.5$\,mm, nearly collinear pieces are merged, $\lmin=1$\,mm\rev{, flights shorter than $\lmin$ are set to $\lmin$}, and $\lmax$ is the longest flight of the condition (larvae) or of the track (termites). The flights then span $L=3$--6 $e$-folds, one to two and a half decades.

\subsection{\emph{Drosophila} larvae}\label{sec:larvae}

Sims \emph{et al.}~\cite{Sims2019} filmed \emph{Drosophila} larvae crawling for up to one hour on plain agar, without odor, light or food, and found L\'evy-like flight statistics that persisted when large parts of the nervous system were switched off. Neurons were targeted with the GAL4--UAS system. The driver BL expresses in the brain lobes and in part of the subesophageal ganglion, BLsens in the same regions and in all sensory neurons, MB247 in the mushroom bodies, a center of learning and memory. The targeted neurons were either silenced with \emph{shibire}$^{ts}$, a temperature-sensitive mutant of dynamin that blocks synaptic transmission above about 30\,$^\circ$C~\cite{Kitamoto2001} (the larvae were tested at 33\,$^\circ$C, some controls also at the permissive 22\,$^\circ$C), or killed by expressing the pro-apoptotic genes \emph{reaper} and \emph{hid} (\emph{rpr,hid}). Larvae that carry only the driver (BL/+, MB247/+) or only the effector (shi/+) are controls. Because the L\'evy statistics survived without brain and senses, Sims \emph{et al.} traced them to the central pattern generator for locomotion in the ventral nerve cord. In an environment controlled so tightly, these data are ideal to ask whether the intrinsic walk of the larva is a renewal process. The 241 tracks with at least 17 flights, 20\,948 flights in ten conditions, are fitted condition by condition with one field per larva and $K\le6$ (Table~\ref{tab:larvae}).

Renewal is rejected in seven of the ten conditions [\figref{fig:walkers}(a)]. The memory is positive, long flights tending to follow long flights, and short, one to four flights deep. It does not depend on how the flights are cut, and in six of the seven conditions it survives when the field of each larva is re-estimated every 25 flights, so it is not a slow drift of the exponent; in BL$>$shi it disappears, and is better described as a slow modulation (Appendix~\ref{app:data}). The nulls are clean: within-larva permutations select $K^*>0$ in 1 of 300 surrogates, and the pipeline null in none of 50. Of the three conditions compatible with renewal, BLsens$>$shi has too few flights (349) to decide, and shi/+ at 22\,$^\circ$C has a lag-1 coupling three standard errors from zero that BIC does not retain; only BL/+ at 22\,$^\circ$C shows no trace of memory. \rev{Temperature matters by itself. The two control genotypes tested at both temperatures have $K^*=0$ at 22\,$^\circ$C and $K^*=1$ at 33\,$^\circ$C, and their lag-1 coupling grows from $-0.004\pm0.019$ to $0.071\pm0.016$ in BL/+, by three standard errors, and from $0.088\pm0.030$ to $0.127\pm0.013$ in shi/+, within errors. BL/+ carries no effector, so heat alone appears to make the walk more persistent, independently of the silencing that it triggers in the lines that express \emph{shibire}$^{ts}$; for those lines, the reference is the controls at 33\,$^\circ$C.}

The memory is strongest in the BLsens$>$\emph{rpr,hid} larvae, whose brain lobes and sensory neurons are dead: $J_1=0.31\pm0.05$, eight standard deviations from the permutation null. In the logic of Ref.~\cite{Sims2019}, the memory between flights is generated below the brain, together with the L\'evy-like statistics, and needs neither the brain nor sensory feedback.

Six of the seven conditions with memory sit at $\Phi\simeq0.7$--1.0 [\figref{fig:overview}(c)], close to the spinodal and about half way to condensation; BL$>$shi, whose memory is a slow modulation, sits at 0.35. Replacing the longest observed flight by the side or the diagonal of the $240\times240$\,mm arena changes $\Phi$ by at most 0.2, because $\mub$ and $J_1$ readjust with $L$. The exponent is visibly dressed [\figref{fig:walkers}(d)]: the one-step rule has $\mub=1.54$--1.77 (2.72 for the ablated larvae\rev{, a value raised by the 28\% of their flights that fall below $\lmin$; Appendix~\ref{app:data}}), the histogram $\hat\mu=1.41$--1.63 (2.52)\rev{. The median observed $\hat\mu$ of every condition is reproduced within 0.035 both by the exact stationary law~\eqref{eq:stationary} of the $K=1$ fits and, for the conditions with $K^*>1$, whose chains have no stationary law in closed form, by simulations of the chains fitted at the selected depth.} Over the full range of flights a truncated log-normal law describes the histogram better than a truncated power law in eight conditions out of ten, so ``L\'evy'' here means at most two decades of approximate scaling; the memory is the same with either single-flight law (Appendix~\ref{app:data}).

\subsection{Termites}\label{sec:termites}

Castiblanco \emph{et al.}~\cite{Castiblanco2022,Castiblanco2019} tracked focal workers of the termite \emph{Cornitermes cumulans} for 6\,h at 2\,Hz, in Petri dishes about 49\,mm across, together with nestmates at different densities. Each of the 75 tracks contains 673 to 12\,091 flights (median 3746), enough to analyze every termite on its own, with $K\le10$.

Every termite remembers. BIC selects a depth between 1 and 8, with a median of 4 [\figref{fig:walkers}(b)], and the lag-1 coupling \rev{of the fit at the selected depth} is positive in all 75 tracks, $J_1=0.34\pm0.10$ (mean $\pm$ s.d.\ over termites); the memory kernel decays over about five flights [\figref{fig:walkers}(c)]. Within-track permutations select memory in 0.3\% of the surrogates, and the pipeline null gives $K^*=0$ in 74 of 75 tracks. With fields re-estimated every 50 or 25 flights the lag-1 coupling stays significant in every termite (median $z=21$ and 18, minimum 6.3).

Termites, like many insects, keep close to walls, a behavior known as thigmotaxis: they spend 30\%, 59\% and 79\% of the time within 3, 6 and 9\,mm of the wall of the dish. The memory could therefore have been geometric, long flights along the wall following each other. It is not. Adding the wall state of each flight as a covariate leaves the lag-1 coupling unchanged (median 0.42 without and 0.41--0.43 with it, for the three bands) and the selected depth unchanged (median 4), and on runs of consecutive flights far from the wall the coupling is even larger, $J_1\simeq0.49$ (Appendix~\ref{app:data}).

The termites sit further into the phase diagram than the larvae, at $\Phi=1.0$--3.5 (median 1.46) [\figref{fig:overview}(c)]: all of them beyond the spinodal, and 7\% beyond the condensation line, consistent with an arena that bounds their longest flights. \rev{These coordinates come from fits with $K=1$, whose lag-1 coupling, 0.42 in the median, absorbs about two thirds of the total coupling of the selected depth, $J_{\rm tot}=0.70$. In the coordinates in which the duality holds at any depth, $\Phi=J_{\rm tot}L/(\mub-1)$ (Sec.~\ref{sec:condensation}), the termites sit deeper, at a median of 1.73 (1.48--2.44), again all beyond the spinodal and 7\% beyond the self-dual line. The cutoff does not move them either: with $\lmax$ set for every termite to the 49\,mm diameter of the dish rather than to its longest flight, $\Phi$ has a median of 1.41, all termites remain beyond the spinodal, and 3\% are beyond the self-dual line.} Their exponents are the most dressed [\figref{fig:walkers}(d)]. The one-step rule has a median $\mub=2.07$, the histogram $\hat\mu=1.49$: read from the histogram, the exponent is off by more than half a unit\rev{, and five termites, beyond the self-dual line, have $\hat\mu<1$ (Sec.~\ref{sec:dressing})}. The apparent exponent grows with the density of nestmates (medians 1.18, 1.51 and 1.59 at densities 0.08, 0.12 and 0.20)\rev{, and it does so through the one-step rule: $\mub$ grows from 1.72 to 2.11 and 2.25, while $\Phi$ changes little, from 1.61 to 1.45 and 1.37. In a more crowded dish termites draw shorter flights, as they would if encounters with nestmates interrupted them, with a memory of nearly unchanged strength; the same trends hold when the flights below $\lmin$ are left out (Appendix~\ref{app:data})}.

\begin{figure*}[t]
\centering
\includegraphics[width=\textwidth]{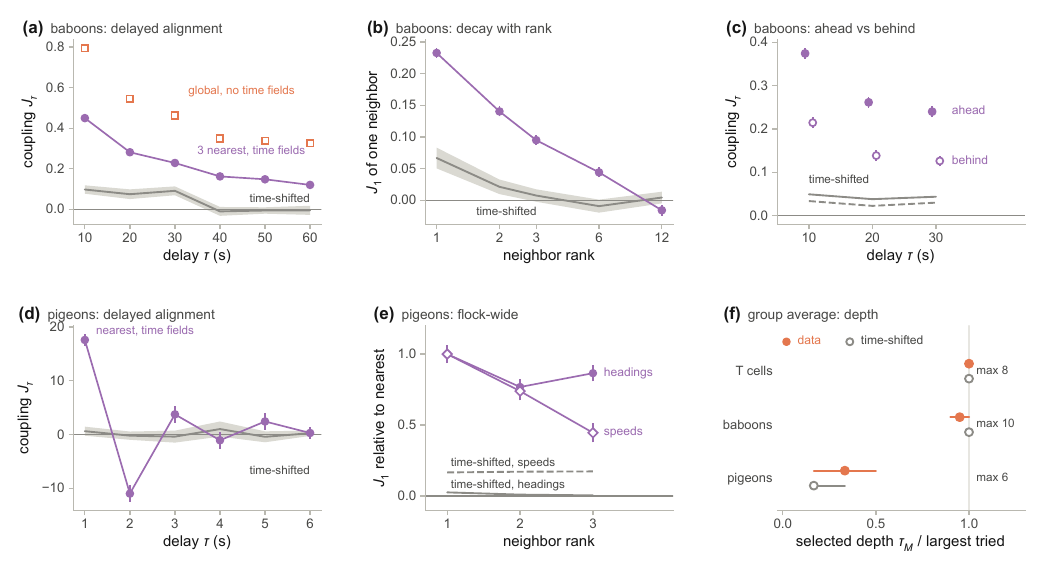}
\caption{Groups. Top: a wild baboon troop~\cite{StrandburgPeshkin2015} (14 days, clock 10\,s, headings). (a)~Couplings of a baboon's heading to the mean heading of its three nearest neighbors at delay $\tau$, with baboon and time fields ($\pm2$ s.e.); grey, mean $\pm2$ s.d.\ over 20 troops in which every baboon is shifted in time; open squares, a global coupling without time fields. (b)~Coupling at 10\,s to a single neighbor of given distance rank, all ranks fitted jointly. (c)~Couplings to the neighbors ahead (filled) and behind (open) along the group heading, among the six nearest; grey, time-shifted troops (solid, ahead; dashed, behind). Bottom: homing-pigeon flocks~\cite{MehlhausenFranks2025} (41 releases, clock 1\,s). (d)~As (a), for the nearest neighbor. (e)~Couplings at 1\,s to a single neighbor of given rank, relative to the nearest, for headings (filled) and log-speeds (open); grey, time-shifted flocks. (f)~Depth $\tau_M$ selected by the group-average analysis, as a fraction of the largest depth tried \rev{(marked on the right, in frames of the clock of each system)}, on the data (coral) and on time-shifted units (grey): median and interquartile range over days (baboons), releases (pigeons) and surrogates; for the T cells, 16 fish pooled.}
\label{fig:groups}
\end{figure*}

\subsection{Baboons and pigeons}\label{sec:realgroups}

A wild troop of olive baboons~\cite{StrandburgPeshkin2015,Crofoot2015data}, 26 animals tracked by GPS every second for 14 days, is a group whose members share a clock and whose positions define neighbors. We use a clock of 10\,s, each day being a co-recorded group ($6\times10^4$ frames in all), and both variables, log-speeds and headings, a heading being defined for steps of at least 1\,m (Appendix~\ref{app:data}). Analyzed through its group average, the troop has a long memory: the selected depth has a median of 9.5 delays over days for headings and is 10, the largest tried, for log-speeds. Time-shifted baboons give the same depths, 10 in 82\% of the day replicates for headings and in all of them for log-speeds [\figref{fig:groups}(f)]. The depth is thus each baboon's own persistence ($\hat a_1=0.70$ for headings, 0.23 for log-speeds), seen through the average as in Eq.~\eqref{eq:aggregation}.

The self\,+\,social model, with baboon and time fields, finds in addition a delayed alignment with the three nearest neighbors, $J_1=0.449\pm0.007$ at 10\,s \rev{(in a troop of this size the time fields inflate heading couplings by about 5\%; Appendix~\ref{app:collective})}, which decays over a minute [\figref{fig:groups}(a)]; on a clock of 30\,s the first coupling is the same, $0.447\pm0.008$. Three controls single it out. Time-shifted baboons keep a coupling of $0.10\pm0.01$ at the shortest delays, because animals that pass the same place at different times head the same way, a field fixed in space that time fields cannot absorb; the data exceed it fourfold. The coupling lives at the nearest neighbors: $0.233\pm0.004$ with the single nearest, then 0.140, 0.095 and 0.044 with the second, third and sixth, and zero with the twelfth [\figref{fig:groups}(b)]. And it is asymmetric: neighbors ahead along the group heading weigh $0.374\pm0.006$, those behind $0.214\pm0.006$, against 0.050 and 0.034 for time-shifted baboons [\figref{fig:groups}(c)]. A baboon thus turns toward where its nearest neighbors, those ahead above all, were heading 10 to 30\,s earlier, in line with the following of initiators documented in this troop~\cite{StrandburgPeshkin2015}; a path walked first by those ahead would act in the same direction, and fields fixed in space would be needed to tell the two apart. Log-speeds tell the same story on a smaller scale, $J_1=0.052\pm0.002$ against $0.017\pm0.001$ for shifted baboons. Read with a global coupling and no time fields, the same headings give $J_1=0.79$ [\figref{fig:groups}(a), squares], the alignment plus the common motion of the troop.

Homing pigeons~\cite{MehlhausenFranks2025,MehlhausenFranks2025data}, three flocks released 41 times, form small groups of three to eight birds per frame, which we follow on a clock of 1\,s. On this scale flight is nearly ballistic, and a bird's heading is extrapolated from its last three, $\hat a_\tau=(185,-82,21)$, that is, from its direction and its rate of turn. \rev{These numbers are large because they are concentrations. Multiplying unit vectors, the couplings add up to the modulus $|\bm\theta|$ of the von Mises law, whose inverse square root is the angular spread of the heading (Sec.~\ref{sec:many}). At 1\,s the heading of a pigeon departs from its extrapolation by a few degrees, which requires $|\bm\theta|$ of order $10^2$; at 10\,s a baboon turns far more, and its couplings are of order one.} The group-average analysis again returns the depths of time-shifted birds, 1 to 3\,s in 80\% of the releases against 91\% for shifted birds [\figref{fig:groups}(f)]. The self\,+\,social model couples each bird to its nearest neighbor with $J_\tau=(17.6\pm0.6,-11.0\pm0.8,3.8\pm0.8)$, against $(0.6\pm0.4,-0.2\pm0.4,-0.4\pm0.6)$ for shifted birds [\figref{fig:groups}(d)]. The alternating signs repeat those of own memory: a bird follows both the last heading of its neighbor and its last turn. \rev{In flocks of three to eight birds the time fields, one per frame, inflate all heading couplings by a common factor, an effect of incidental parameters~\cite{NeymanScott1948} that on synthetic flocks with these couplings amounts to 1.9, 1.5 and 1.3 for three, five and eight birds (Appendix~\ref{app:collective}). The values above are therefore too large in scale, but not in sign, in their ratios or in their comparison with the time-shifted birds, which share flocks and fields; log-speed couplings show no such inflation.} In flocks this small the heading coupling does not decay over the three nearest neighbors, whereas speeds match the nearest one most [\figref{fig:groups}(e)]; log-speeds are coupled with $J_1=3.8\pm0.1$, against $0.5\pm0.2$ for shifted birds. Nor is the coupling stronger toward birds ahead ($3.5\pm0.3$ ahead, $5.3\pm0.3$ behind among the three nearest): pigeons follow their leaders with delays of fractions of a second~\cite{Nagy2010}, within one frame of this clock, and a model of delayed couplings sees only the interactions slower than its frame.

\subsection{A negative control: T cells}\label{sec:tcells}

T cells close the analysis as a negative control, first for single walkers. Jerison and Quake~\cite{Jerison2020} tracked T cells, a class of leukocytes, as they migrate in vivo through the tissue of living zebrafish, and showed that the apparent heavy tails of the steps of these cells come from differences between cells, so any memory found with one field for all cells should be heterogeneity in disguise. It is. With one field for all cells, every group of T cells (and of \emph{Dictyostelium} amoebae from the same data set) shows deep memory, and it still does after the steps of each cell are shuffled ($K^*=5$--6 in 49 of 50 surrogates), which can only be heterogeneity. With one field per cell the shuffled data give $K^*=0$ in every surrogate, the memory of mouse T cells disappears, and zebrafish T cells and \emph{Dictyostelium} keep a short persistence ($J_1\simeq0.14$--0.15 per frame; Appendix~\ref{app:data}).

The same cells also test the group analysis. Zebrafish T cells co-recorded within each fish on a common clock of 45\,s (16 fish, 712 cells, 85\,016 steps) remember their own speed for about four frames ($\hat a_1\simeq0.20$). The group-average analysis selects a delay of 8 frames, the largest tried, and it selects the same delay in 20 of 20 time-shifted surrogates [\figref{fig:groups}(f)]: the delay is own memory. A weak coupling to the three nearest cells, $J_1=0.021\pm0.006$, vanishes once time fields are included, $0.005\pm0.007$; what remains is a drive shared within each fish.

These findings characterize the relevant timescale of speed correlations rather than ruling out immune interactions broadly. While immune cells interact biochemically, synaptically, and collectively during target swarming~\cite{Agliari2014,Sabatos2008,Gerard2013,GaleanoNino2020,Agliari2020}, in the baseline tissue conditions of Ref.~\cite{Jerison2020} these interactions do not manifest as delayed speed correlations on a 45\,s clock. The zebrafish T cells therefore provide a clean empirical negative control: a benchmark collective where apparent delays detected by group averaging stem entirely from individual cellular persistence.

The same model and the same controls thus return different answers for different groups: a delayed alignment of baboons with their nearest neighbors, stronger toward those ahead; a flock-wide coupling between pigeons, whose leader--follower delays are shorter than a frame; and no delayed coupling between the speeds of T cells. In none of the three could the answer be read off the group average, whose selected depth is in each case the own memory of the units.

\section{Discussion}\label{sec:discussion}

L\'evy walks are usually judged by the histogram of their flights, while the renewal hypothesis on which the textbook model rests, that every flight is drawn afresh, is seldom tested directly. To our knowledge, this work is one of the first to test it with a model of the sequence of flights whose depth the data select, and the first to do so with maximum entropy. The renewal L\'evy walk turns out to be a convenient idealization, and most of the walkers analyzed here do not follow it. Larvae and termites remember their recent flights, positively and over a few flights, and the memory survives every control we could think of: one field per individual, fields that drift along the track, other ways of cutting tracks into flights, permutation and pipeline nulls, and, for the termites, the wall that they follow. In the larvae it is strongest when brain lobes and sensory neurons have been ablated, which places its origin, together with that of the L\'evy-like statistics~\cite{Sims2019}, below the brain. Part of the dependence between flights may come from switches between behavioral states, which a chain of flights describes as memory; couplings within each state would apportion the two~\cite{Stoye2025}.

Memory changes what a histogram means. A walker that packs long flights into runs has a heavier-tailed histogram than the rule it follows at each turn, by more than half a unit of exponent in the termites. An exponent read from a histogram, often compared with the value $\mu=2$ of optimal search~\cite{Viswanathan1999}, need not be the exponent of the rule that generates the walk. Memory also gives the cutoffs a physical role: the coupling acts through $JL$, the flights condense at the largest scale beyond a self-dual coupling, and over the one or two decades accessible to experiments the walk feels that transition well before reaching it: every termite has crossed its spinodal. Two extensions follow directly: a joint model of the length and the direction of flights, with couplings between speed and turning~\cite{Jerison2020}, and the transport properties of L\'evy walks whose flights are coupled in this way, to which renewal theory does not apply\rev{, and in which the time spent in runs of long flights condenses before their number does (Sec.~\ref{sec:condensation})}.

In groups, the same formalism separates what the group average mixes: each unit's memory of its own past, drives common to all units, and the influence of the others. Read through the average, own memory appears as a delayed coupling whose strength does not decrease with the size of the group, and a common drive as a coupling at any delay. Read with the self\,+\,social model, the delays found in zebrafish T cells are own memory and a drive shared within each fish, baboons align with where their nearest neighbors, those ahead above all, were heading 10 to 30\,s earlier, and pigeons are coupled to their whole flock from one second to the next. The recipe applies to any attempt to infer delayed interactions from trajectories: model units, not their average; give each unit its own field and memory and each frame its own field; couple units to their neighbors rather than to the group; compare with time-shifted units; and check how a coupling depends on the distance and on the position of the neighbor. Fields fixed in space, such as trails or a homing route, survive both time fields and time shifts, and are the natural next ingredient. Finally, the collective speed has a critical point of its own, at $J_c=12/L^2$, set by the largest susceptibility of a single flight: a delayed positive coupling between scale-free speeds produces collective stop-and-go, a negative one oscillations, and own memory lowers the threshold.

\section*{Data and code availability}
All trajectories analyzed in this work are publicly available~\cite{Sims2019,Castiblanco2019,Jerison2020,Crofoot2015data,MehlhausenFranks2025data}. All code to reproduce the analyses and figures is openly available at \url{https://github.com/coscienza-research/levy-walks-maxent}.

\begin{acknowledgments}
We are grateful to the authors of Refs.~\cite{Sims2019,Castiblanco2022,Jerison2020,StrandburgPeshkin2015,MehlhausenFranks2025} for making their data public. 
A.B., A.L. acknowledge financial support from the \rev{Italian} National Institute for Higher Mathematics (INdAM) through the GNFM group. Additionally, A.B.
acknowledges support from INFN, Sezione di Roma 1.
A.B.\ acknowledges the grant NEEDLE from Sapienza Universit\`a di Roma. A.L.\ acknowledges support from the project {ULTRACLOCK} from Sapienza Universit\`a di Roma.
The authors acknowledge the use of the Lagrange Multi-GPU Server at the Department of Mathematics, Sapienza University of Rome, for computational resources supporting this work.
\end{acknowledgments}

\appendix
\section{One flight and the chain with one lag}\label{app:single}

\begin{figure*}[t]
\centering
\includegraphics[width=\textwidth]{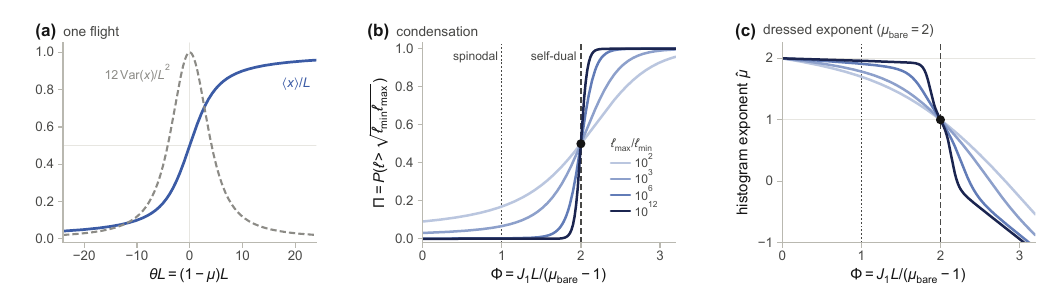}
\caption{The memory chain at depth one. (a)~Response of one flight to its field, Eq.~\eqref{eq:langevin}: mean (solid) and variance (dashed, in units of its maximum $L^2/12$) of $x$. (b)~Order parameter $\Pi$, the fraction of stationary flights longer than $\sqrt{\lmin\lmax}$, versus $\Phi=J_1L/(\mub-1)$ at $\mub=2$, for ranges $\lmax/\lmin$ from $10^2$ to $10^{12}$. The curves cross at the self-dual point $\Phi=2$, $\Pi=1/2$, and sharpen into a step as the range grows; dotted, the spinodal $\Phi=1$. (c)~Exponent returned by an i.i.d.\ truncated-Pareto fit to the same stationary laws: memory lowers it below $\mub$, and every curve crosses $\hat\mu=1$ at $\Phi=2$.}
\label{fig:theory}
\end{figure*}

\paragraph{Single-site law.} For $p(x)\propto e^{\theta x}$ on $[0,L]$, $Z(\theta)=(e^{\theta L}-1)/\theta$ and $\ln Z(\theta)=\theta L/2+\ln[2\sinh(\theta L/2)/\theta]$, so that
\begin{equation}
\frac{d\ln Z}{d\theta}=\frac L2+\frac L2\coth\frac{\theta L}2-\frac1\theta=\Lambda(\theta),\qquad \frac{d^2\ln Z}{d\theta^2}=\Lambda'(\theta),
\end{equation}
which is Eq.~\eqref{eq:langevin} [\figref{fig:theory}(a)]. With $u=2x/L-1$, $p(u)\propto e^{bu}$ on $[-1,1]$ with $b=\theta L/2$, and $\langle u\rangle=\Lang(b)$: this is the law of $\cos\vartheta$ for a classical unit vector with Boltzmann weight $e^{b\cos\vartheta}$, since the area element $\sin\vartheta\,d\vartheta\,d\varphi=-d(\cos\vartheta)\,d\varphi$ is uniform in $\cos\vartheta$. The maximum of $\Lambda'$ is at $\theta=0$, where $\Lang'(0)=1/3$ gives $\Lambda'(0)=L^2/12$. Adding the constraint $\langle x^2\rangle$ gives $p(x)\propto e^{\theta x-\beta x^2}$, a log-normal law truncated to $[\lmin,\lmax]$, which we use as a nested alternative, with integrals computed by Gauss--Legendre quadrature.

\paragraph{Detailed balance.} With $\pi(x)\propto e^{hx}Z(h+Jx)$ and the kernel $p(y|x)=e^{(h+Jx)y}/Z(h+Jx)$,
\begin{equation}
\pi(x)\,p(y|x)\propto e^{hx+(h+Jx)y}=e^{h(x+y)+Jxy},
\end{equation}
which is symmetric in $(x,y)$; $\pi$ is therefore stationary and the chain reversible. Under $x\to L-x$, $y\to L-y$ the exponent becomes $2hL+JL^2-(h+JL)(x+y)+Jxy$, the same weight with $h\to-h-JL$; the self-dual line is $h=-JL/2$.

\paragraph{Path versus causal measure.} The path measure~\eqref{eq:joint} with $K=1$ is a reversible Markov chain whose kernel is the Doob transform of its transfer operator, $p_{\rm path}(y|x)=e^{hy+Jxy}\varphi(y)/[\lambda_0\varphi(x)]$, with $\varphi$ the Perron eigenfunction of the kernel $e^{h(x+y)/2+Jxy}$ on $[0,L]$ and $\lambda_0$ its eigenvalue. The causal kernel is the same expression with $\varphi\equiv1$ and normalization $Z(h+Jx)$. Since $\ln\varphi(y)=J\langle x\rangle_0\,y+O(J^2)$, with $\langle x\rangle_0$ the mean at $J=0$, the two families coincide to first order in $J$ after the shift $h_{\rm causal}=h_{\rm path}+J\langle x\rangle_0$, and they differ at order $J^2$ through an effective log-normal curvature that grows near the thresholds. Fitted to $T=6000$ flights generated by either family, the two return $\hat J=0.095$ and 0.099, with log-likelihoods within 2 nats.

The path measure has the same duality, with self-dual line $h=-JL$ instead of $h=-JL/2$. In the path measure each flight interacts with both its neighbors in time, so a run of flights at the cutoff exerts on a flight a field $h+2JL$, against $h+JL$ in the causal chain, and the self-dual coupling is half as large at the same bare field. This difference in the self-dual couplings reflects the shift in the baseline field: because $h_{\rm causal}$ and $h_{\rm path}$ differ by $J\langle x\rangle_0$, matching nominal fields do not correspond to the same walk. Throughout this work, we focus exclusively on the causal formulation. We verified numerically, with the transfer operator on 160 Gauss--Legendre nodes, that on the self-dual line of the path measure $\Pi=1/2$ for every $L$, and that the crossover sharpens with $L$ as for the causal chain.

\section{Large-range theory of the condensation}\label{app:largeL}

For $\theta L\to+\infty$, $\ln Z(\theta)=\theta L-\ln\theta+o(1)$; for $\theta L\to-\infty$, $\ln Z(\theta)=-\ln|\theta|+o(1)$. With $h=-\alpha$, $J=\beta/L$ and $x=uL$, the field after a flight $x$ is $\theta=-\alpha+\beta u$, and
\begin{equation}
\ln\pi(uL)=-\alpha uL+L\max(\beta u-\alpha,0)+O(\ln L),
\end{equation}
which is Eq.~\eqref{eq:freeenergy}. $F(u)$ is piecewise linear: it increases with slope $\alpha$ for $u<\alpha/\beta$ and, if $\beta>\alpha$, decreases with slope $\alpha-\beta$ beyond. The end $u=1$ is thus a local minimum for $\beta>\alpha$ (spinodal, $J=\Jup$) and the global one for $\beta>2\alpha$ (coexistence, $J=\Jcond$), where $F(1)=2\alpha-\beta$ changes sign. The mass near each minimum is $e^{-LF}$ up to algebraic prefactors, so $\Pi\to\Theta(\beta-2\alpha)$ as $L\to\infty$, a first-order jump. At finite $L$ the exact condition for a secondary maximum of $\pi$ at the cutoff is $J\,\Lambda(h+JL)>\mub-1$, which tends to $J>\Jup$ for wide ranges; below $\Jcond$ its weight is exponentially small, $\pi(L)/\pi(0)\sim e^{(2h+JL)L}$.

Metastability can be seen directly. Starting a chain at $x=L$ in the window $\alpha<\beta<2\alpha$, the first flight shorter than $\sqrt{\lmin\lmax}$ occurs after a mean number of flights that grows exponentially with $L$: at $\beta=1.8\alpha$, from 4 at $\lmax/\lmin=10^2$ to 9 at $10^4$ and 73 at $10^8$ (4000 simulated chains each). In this window a walker that enters a run of long flights stays in it for a time set by the range, the dynamical signature of a metastable phase.

\section{Estimation}\label{app:estimation}

\paragraph{Likelihood and Newton's method.} Writing $\theta_{i,t}=\bm w\cdot\bm z_{i,t}$, with $\bm w$ collecting fields and couplings and $\bm z_{i,t}$ the corresponding regressors (indicators of the walker or block, lagged flights, covariates), the negative log-likelihood $-l=\sum_{i,t}[\ln Z(\theta_{i,t})-\theta_{i,t}x_{i,t}]$ has gradient $\sum[\Lambda(\theta)-x]\bm z$ and Hessian $\sum\Lambda'(\theta)\,\bm z\bm z^{\mathsf T}$, positive semi-definite. We use Newton's method with a backtracking line search, started from the renewal solution, which converges in a few iterations to a gradient below $10^{-8}$ per flight. The Hessian at the optimum is the observed Fisher information~\eqref{eq:fisher}; the sandwich covariance $\mathsf I^{-1}\mathsf G\,\mathsf I^{-1}$, with $\mathsf G$ the sum of outer products of the per-flight scores, guards against misspecification and is used for the couplings of the fits with one field per walker. All depths $K\le K_{\max}$ are fitted on the same flights, those with at least $K_{\max}$ predecessors, so that they can be compared. The timings quoted in Sec.~\ref{sec:estimation} use a dense Hessian on a single thread.

\paragraph{Centering and the bare exponent.} In the fits the lags are centered, $\theta_t=h+\sum_\tau J_\tau(x_{t-\tau}-c)$, with $c$ the mean of $x$ (of each track, with one field per walker), which decorrelates the field from the couplings. The bare exponent is $\mub=1-h'$, with $h'=h-c\sum_\tau J_\tau$ the field after a run of the shortest flights; its standard error follows from the delta method.

\paragraph{Fields per walker, local fields and covariates.} With one field per walker, $\bm z_{i,t}$ contains the indicator of walker $i$; with local fields, the indicator of the block of $B$ flights to which $t$ belongs (the last block is merged with the previous one if shorter than $B/2$); a known covariate $w_{i,t}$, such as the wall state of a flight, adds a column, $\theta_t\to\theta_t+g\,w_t$. The incidental-parameter bias of the couplings, of order $1/T_i$~\cite{Nickell1981}, is measured on surrogates in which the flights of each walker (or of each block) are permuted, and the $z$-scores of Table~\ref{tab:larvae} are computed against these permutation distributions.

\section{Model selection and validation}\label{app:selection}

\begin{figure*}[t]
\centering
\includegraphics[width=\textwidth]{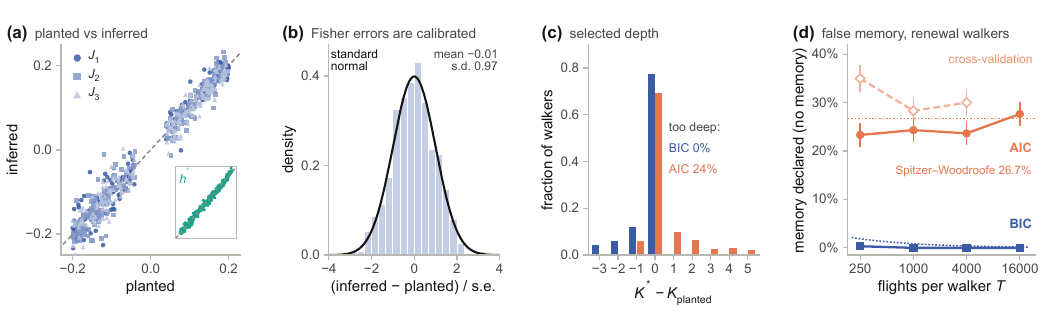}
\caption{Validation on planted walkers. (a)~Planted versus inferred couplings $J_1$, $J_2$, $J_3$ for 200 walkers of $T=3000$ flights, with bare exponents drawn uniformly in $[1.3,2.5]$, couplings of random sign and modulus in $[0.04,0.20]$, and $\lmax/\lmin=10^2$; inset: the field $h'=1-\mub$. (b)~Errors in units of the Fisher standard errors, pooled over all parameters, against the standard normal law. (c)~Selected depth minus planted depth: BIC never overestimates it, and underestimates it only when the last coupling is weaker than about three standard errors; AIC overestimates it in a quarter of the walkers. (d)~Fraction of renewal walkers ($\mub=2$, $\lmax/\lmin=10^3$, 300 replicates per length) declared to have memory by AIC, blocked cross-validation and BIC; dotted lines, the asymptotic values of Eq.~\eqref{eq:woodroofe} for AIC and $P(\chi^2_1>\ln T)$ for BIC.}
\label{fig:validation}
\end{figure*}

\paragraph{AIC and BIC under the null.} For nested models of depths $0,\dots,K_{\max}$, under the null the likelihood-ratio statistics of successive increments are asymptotically independent $\chi^2_1$ variables, so that $\Delta_k=\sum_{j\le k}(\chi^2_{1,j}-2)$ is a random walk with negative drift, and AIC selects the depth at which it is largest. Spitzer's combinatorial lemma gives the probability that the maximum is attained at $k=0$, hence Eq.~\eqref{eq:woodroofe}~\cite{Shibata1976,Woodroofe1982}. Replacing the penalty 2 by $\ln n_{\rm o}$ (BIC) makes the drift diverge, and the false-positive rate decreases as $P(\chi^2_1>\ln n_{\rm o})$ to leading order. Cross-validation with a fixed training fraction estimates the same expected out-of-sample log-likelihood as AIC and is asymptotically equivalent to it~\cite{Stone1977}; it becomes consistent only if the training fraction vanishes~\cite{Shao1993}. Our blocked cross-validation splits each track into five contiguous folds separated by gaps of $K_{\max}$ flights.

\paragraph{Planted walkers.} Walkers generated by the causal chain with random parameters (\figref{fig:validation}) lie along the diagonal of planted versus inferred couplings, and the errors in units of the Fisher standard errors have mean $-0.01$ and standard deviation 0.97. BIC recovers the planted depth in 78\% of the walkers and never selects a deeper one; when it errs, the last coupling is below detection. AIC selects a deeper chain in 24\% of the walkers. On renewal walkers the false-positive rate of AIC stays at the level of Eq.~\eqref{eq:woodroofe} for every length from 250 to 16\,000 flights, that of blocked cross-validation is slightly higher (28--35\%), and that of BIC vanishes as $P(\chi^2_1>\ln T)$. The rates are the same, 21--25\% for AIC and at most 0.5\% for BIC, for exponents $\mu=1.5$--2.5 and for ranges up to five decades.

\section{What can look like memory}\label{app:pitfalls}

\begin{figure*}[t]
\centering
\includegraphics[width=\textwidth]{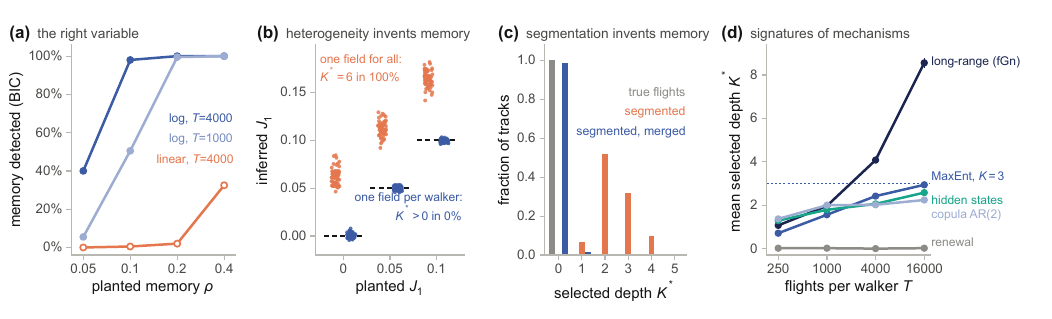}
\caption{How memory is missed or invented, on synthetic walkers. (a)~Power to detect a weak memory with the log statistic $x_tx_{t-1}$ and with the linear statistic $\ell_t\ell_{t-1}$ (BIC; 200 replicates per point). (b)~Inferred lag-1 coupling for 40 walkers with individual exponents $\mu_i\sim U(1.2,2.2)$ and $T=1000$ flights each, with a common field (coral) or one field per walker (indigo); dashed, planted values. (c)~Selected depth for a renewal L\'evy walk sampled at fixed times with localization noise: true flights, flights segmented from the track, and segmented flights after merging nearly collinear pieces. (d)~Mean selected depth ($\pm$ s.e.) versus the number of flights for five mechanisms with truncated-Pareto marginals: renewal, a MaxEnt chain of depth 3 (dotted: its true depth), a Gaussian-copula AR(2) process, a walker switching between two hidden states, and long-range memory (Gaussian copula of fractional Gaussian noise with Hurst exponent 0.75).}
\label{fig:pitfalls}
\end{figure*}

\paragraph{The variable.} The statistic $\ell_t\ell_{t-1}$, the natural guess in linear space, is dominated by a handful of extreme flights. For a weak memory, a Gaussian-copula autoregression with truncated-Pareto marginals and correlation $\rho=0.1$ ($T=4000$), the log statistic $x_tx_{t-1}$ detects it in 98\% of the walkers and the linear one in none [\figref{fig:pitfalls}(a)].

\paragraph{Heterogeneity.} With 40 memoryless walkers whose exponents are spread over $[1.2,2.2]$, a model with a common field selects the maximal depth in every replicate and returns $J_1\simeq0.11$ for a planted $0.05$. One field per walker removes both the false positives and the bias [\figref{fig:pitfalls}(b)].

\paragraph{Corridor segmentation.} Positions are sampled at fixed times with noise, and flights must be cut out of the track. A flight is the longest stretch of track that stays within a corridor of half-width $\varepsilon$ around the segment joining its end points~\cite{Rhee2011,Humphries2013}; the track is scanned from its start, each flight being extended until a point leaves the corridor, and its length is the distance between its end points.

\paragraph{Merging.} Localization noise displaces single points out of the corridor and splits a long flight into two long pieces, a positive correlation between consecutive lengths that the analysis reads as memory: a renewal L\'evy walk segmented in this way shows $K^*=2$--3 [\figref{fig:pitfalls}(c)], and the lag-1 autocorrelation of $\ln\ell$ rises from 0.02 to 0.21. We merge consecutive pieces whose headings differ by less than $\theta_c=20^\circ$, which restores $K^*=0$ while leaving genuine memory detectable. Applied to true flights with independent turning angles, merging deletes boundaries independently of the lengths, an i.i.d.\ thinning of the renewal process, which therefore remains renewal; applied to noise-split pieces, it restores the original flights. The test of \figref{fig:pitfalls}(c) uses renewal L\'evy walks of 3000 flights with $\mu=2$ over three decades, uniformly random directions and unit speed, sampled at $\Delta t=0.25$ with Gaussian localization noise of standard deviation 0.02, and segmented with $\varepsilon=0.1$ and $\lmin=1$.

\paragraph{Pipeline null.} On real data the track is cut at its flight boundaries; the pieces, each with its own sub-sampled points, are reshuffled, rotated, and reconnected with turning angles drawn from the empirical distribution; the whole pipeline (segmentation, merging, fit) is then run again on the rebuilt track. Any memory found there is produced by the pipeline.

\paragraph{Signatures of mechanisms.} Mechanisms of dependence differ in how the selected depth grows with the length of the data [\figref{fig:pitfalls}(d)]. $K^*$ stays at zero for a renewal walk and saturates at the true order for a finite memory. It grows slowly for a walker that switches between hidden states, whose dependence decays geometrically and becomes resolvable lag after lag, and without bound for long-range memory.

\section{Data sets and controls on single walkers}\label{app:data}

\begin{table}[b]
\caption{Data sets.}
\label{tab:datasets}
\begin{tabular}{ll}
\hline\hline
system & content and source\\
\hline
\emph{Drosophila} larvae & \parbox[t]{6.2cm}{304 tracks at 2\,Hz, up to 1\,h, 10 genotype--temperature conditions, $240\times240$\,mm agar arenas~\cite{Sims2019}; Dryad doi:\allowbreak 10.5061/\allowbreak dryad.7m0cfxpq0, mirror Zenodo 4996872 (CC0)}\\[1ex]
termite workers & \parbox[t]{6.2cm}{75 focal \emph{C.~cumulans} workers, 2\,Hz for 6\,h, Petri dishes of $\approx49$\,mm~\cite{Castiblanco2019}; Zenodo 3628061 (CC-BY 4.0)}\\[1ex]
T cells & \parbox[t]{6.2cm}{1106 tracks of zebrafish and mouse T cells and \emph{Dictyostelium}~\cite{Jerison2020}; GitHub erjerison/TCellMigration (MIT)}\\[1ex]
baboons & \parbox[t]{6.2cm}{26 GPS-collared members of a troop of olive baboons (\emph{Papio anubis}), 1\,Hz from 06:00 to 18:00 on 14 days~\cite{StrandburgPeshkin2015}; Movebank Data Repository doi:\allowbreak 10.5441/\allowbreak 001/1.kn0816jn (CC0)~\cite{Crofoot2015data}}\\[1ex]
pigeons & \parbox[t]{6.2cm}{three flocks of homing pigeons (6, 10 and 8 tracked birds), 41 releases, GPS at 1\,Hz (5\,Hz for some birds)~\cite{MehlhausenFranks2025}; Dryad doi:\allowbreak 10.5061/\allowbreak dryad.4tmpg4fp6 (CC0)~\cite{MehlhausenFranks2025data}}\\
\hline\hline
\end{tabular}
\end{table}

\begin{table*}[t]
\caption{Memory between consecutive flights of \emph{Drosophila} larvae. BL drives expression in the brain lobes and part of the subesophageal ganglion, BLsens also in all sensory neurons, MB247 in the mushroom bodies; ``shi'' is the temperature-sensitive \emph{shibire}$^{ts}$, which silences the targeted neurons at 33\,$^\circ$C, and \emph{rpr,hid} the pro-apoptotic genes that kill them; X/+ and shi/+ carry a single transgene. $K^*$: depth selected by BIC with one field per larva. $J_1$: lag-1 coupling of the fit with $K=1$ ($\pm$ s.e.). $z_{\rm sh}$: $J_1$ against 30 within-larva permutations of the flights. $z_{25}$: the same with fields re-estimated every 25 flights, against permutations within blocks (truncated power law / truncated log-normal single-flight law). $\Phi=J_1L/(\mub-1)$: median over larvae. The permutation null selects $K^*>0$ in 1 of 300 surrogates, the pipeline null in none of 50.}
\label{tab:larvae}
\begin{ruledtabular}
\begin{tabular}{llrrcr@{\,$\pm$\,}lrrr}
condition & manipulation & larvae & flights & $K^*$ & \multicolumn{2}{c}{$J_1$} & $z_{\rm sh}$ & $z_{25}$ & $\Phi$\\
\colrule
BL/+, 22\,$^\circ$C & control & 24 & 1976 & 0 & $-0.004$ & 0.019 & 0.2 & 0.3 / 0.1 & 0.07\\
shi/+, 22\,$^\circ$C & control & 19 & 1366 & 0 & 0.088 & 0.030 & 3.2 & 3.3 / 3.5 & 0.56\\
\emph{rpr,hid} control & control & 59 & 4240 & 1 & 0.074 & 0.012 & 12.3 & 6.6 / 8.2 & 0.77\\
BL/+, 33\,$^\circ$C & control & 26 & 2337 & 1 & 0.071 & 0.016 & 5.3 & 3.6 / 4.8 & 0.69\\
shi/+, 33\,$^\circ$C & control & 29 & 2605 & 1 & 0.127 & 0.013 & 11.2 & 10.4 / 7.4 & 1.02\\
MB247/+ & control & 21 & 1630 & 2 & 0.139 & 0.021 & 7.7 & 6.4 / 6.5 & 0.86\\
MB247$>$shi & mushroom bodies silenced & 22 & 2103 & 3 & 0.168 & 0.021 & 9.7 & 8.3 / 6.7 & 1.04\\
BL$>$shi & brain lobes silenced & 17 & 2366 & 4 & 0.058 & 0.025 & 2.8 & $-2.1$ / $-2.3$ & 0.35\\
BLsens$>$shi & brain lobes and sensory neurons silenced & 6 & 349 & 0 & $-0.063$ & 0.106 & $-0.2$ & $-2.2$ / $-1.9$ & $-0.32$\\
BLsens$>$\emph{rpr,hid} & brain lobes and sensory neurons ablated & 18 & 1976 & 2 & 0.307 & 0.045 & 8.2 & 5.0 / 4.0 & 0.72\\
\end{tabular}
\end{ruledtabular}
\end{table*}

\begin{figure*}[t]
\centering
\includegraphics[width=\textwidth]{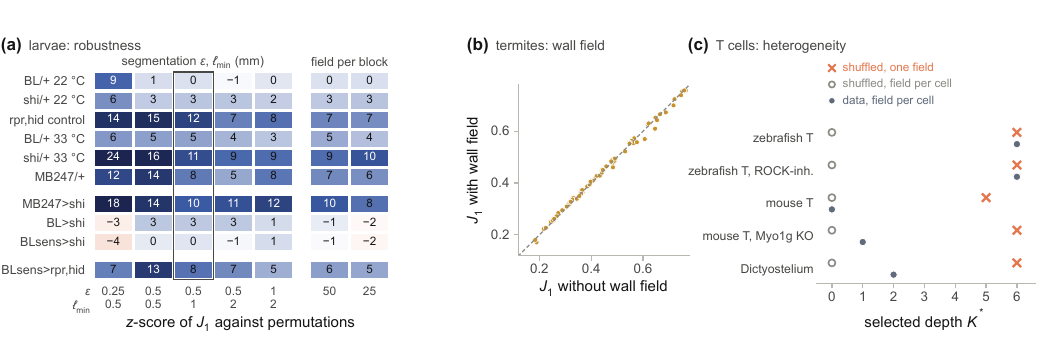}
\caption{Controls on real walkers. (a)~\emph{Drosophila} larvae: $z$-score of $J_1$ against permutation nulls for five segmentation settings (corridor half-width $\varepsilon$ and $\lmin$, in mm; boxed, the setting of Table~\ref{tab:larvae}), and with fields re-estimated every 50 or 25 flights, against permutations within blocks. (b)~Termites: lag-1 coupling of a $K=1$ fit with and without a wall field (band of 6\,mm), one point per termite. (c)~T cells and \emph{Dictyostelium}~\cite{Jerison2020}: depth selected after shuffling the steps within each cell, with a common field (crosses) or one field per cell (open circles), and on the real data with one field per cell (filled).}
\label{fig:controls}
\end{figure*}

\paragraph{Preprocessing.} Table~\ref{tab:datasets} lists the data. For the larvae \rev{and the termites}, flights below $\lmin=1$\,mm (5--13\% of all flights \rev{in the larval conditions}, 28\% for the two lines with silenced or ablated sensory neurons\rev{, and a median of 15\%, 2--30\%, of the flights of a termite}) are set to $\lmin$\rev{. For the larvae}, $\lmax$ is the longest flight of the condition. For the termites, $\lmax$ is the longest flight of each track (26--45\,mm), and the wall state of a flight is the fraction of its points within 3, 6 or 9\,mm of the wall, the arena being estimated from the track as the circle that contains 99.5\% of its points. For the T cells the variable is the displacement per frame, with $\lmin$ at the 5th percentile. For the collective analyses, $\lmin$ is at the localization floor (5th percentile) and steps above the 99.95th percentile are treated as tracking glitches. Baboon positions are taken every 10\,s (every 30\,s for the robustness check) and pigeon positions every second, keeping the first fix of each second for the birds logged at 5\,Hz. A heading is defined for steps of at least 1\,m (baboons, about four times the GPS error) or 3\,m (pigeons); a unit that moves less contributes a zero vector to its own lags and to its neighbors' regressors\rev{, and its heading at that frame, being undefined, is left out of the likelihood}. Each day (baboons) or release (pigeons) is a co-recorded group, with its own unit and time fields.

\paragraph{Larvae.} Table~\ref{tab:larvae} details the ten conditions. The memory does not depend on the segmentation, with $\varepsilon\in\{0.25,0.5,1\}$\,mm and $\lmin\in\{0.5,1,2\}$\,mm, and in six of the seven conditions with memory it survives fields re-estimated every 50 or 25 flights [\figref{fig:controls}(a)]; in BL$>$shi it does not. Over the full range $[\lmin,\lmax]$ a truncated log-normal law, the MaxEnt law with $\langle x^2\rangle$ also fixed, describes the histogram better than a truncated power law in eight conditions out of ten ($\Delta$BIC$\,=286$--1625); the power law is competitive only for the larvae with silenced or ablated sensory neurons. This does not contradict tail fits above a data-chosen $x_{\min}$~\cite{Clauset2009}, but ``L\'evy'' here refers to two decades at most. The memory is the same with either single-flight law (Table~\ref{tab:larvae}, $z_{25}$).

\paragraph{\rev{Flights below $\lmin$.}} \rev{Setting the flights below $\lmin$ to $\lmin$ places a point mass at $x=0$, which the continuous law~\eqref{eq:single} does not have, and removes the memory term from the field of the next flight. The alternative is to condition on $\ell\ge\lmin$. The law $e^{\theta x}/Z(\theta)$ on $[0,L]$ is exactly that of a flight conditioned to exceed $\lmin$, so the flights below $\lmin$ can be dropped from the likelihood as responses and kept, at $x=0$, among the lags. In the seven larval conditions with memory this lowers $J_1$ by 23--41\%. The coupling stays positive in all of them, and at least 2.5 standard errors from zero in all but BL$>$shi, whose memory is a slow modulation. The bare exponent drops as well, most in the ablated larvae, from 2.72 to 2.09; their coupling, $J_1=0.19\pm0.04$, remains the largest of all conditions. Since both enter $\Phi=J_1L/(\mub-1)$, $\Phi$ moves by at most 0.2. In the termites, conditioning lowers the median $J_1$ of the $K=1$ fits from 0.42 to 0.28 and the median $\mub$ from 2.07 to 1.67, and leaves every termite beyond the spinodal (median $\Phi=1.61$); $\mub$ still grows with the density of nestmates (medians 1.46, 1.68 and 1.75) while $\Phi$ decreases slightly (1.78, 1.58 and 1.48). Dropping also the flights that follow one below $\lmin$, or giving them a field of their own, changes the larval couplings of the conditions with memory by less than 0.025.}

\paragraph{Termites and the wall.} With the wall state of each flight as a covariate, the lag-1 coupling is unchanged in every termite [\figref{fig:controls}(b)]. The wall field improves BIC by a few units for the typical termite (median $-3.5$ at 6\,mm) and by much more for a few, without affecting their memory.

\paragraph{T cells.} With a common field, shuffled T cells select $K^*=5$--6 in 49 of 50 surrogates, the maximal depth in 43; with one field per cell, $K^*=0$ in every surrogate [\figref{fig:controls}(c)]. The steps are log-normal rather than power-law distributed, by $10^2$--$10^4$ BIC units.

\section{The collective model}\label{app:collective}

\begin{figure*}[t]
\centering
\includegraphics[width=\textwidth]{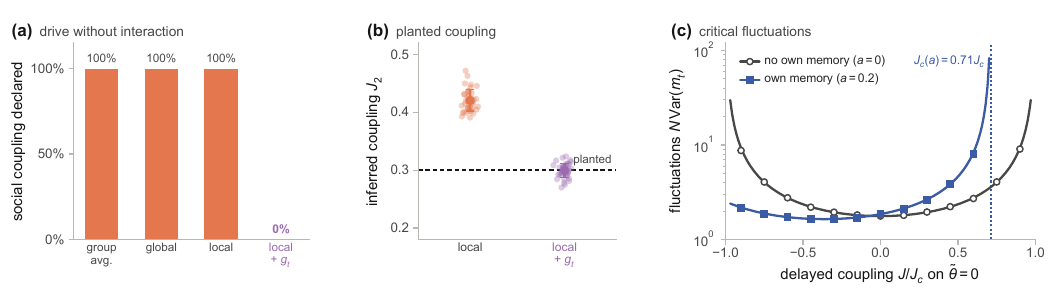}
\caption{Synthetic collectives. (a)~A drive common to all units, without interaction (40 replicates): fraction of data sets in which a social coupling is declared by the group-average analysis, by a global coupling, by a local coupling, and by a local coupling with time fields; right, a planted local coupling $J_2=0.3$ under the same drive, inferred without and with time fields (mean $\pm$ s.d.). (b)~Fluctuations of the collective log-speed versus $J$ on the line $\tilde\theta=0$, without (open circles) and with own memory $a=0.2$ (squares); lines, the theory of Sec.~\ref{sec:meanfield}; dotted, the shifted threshold $J_c(a)=0.71J_c$. $L=\ln100$.}
\label{fig:collective}
\end{figure*}

\begin{table*}[t]
\caption{Synthetic collectives: the analysis of the group average of Ref.~\cite{Agliari2024} (depth $\tau_M^*$ selected by AIC, delay given by $\arg\max_\tau|J_\tau|$) against the self\,+\,social model~\eqref{eq:selfsocial} (depths by BIC). ``Declared'': a nonzero social coupling is selected. Speed-Vicsek model, $L=\ln100$, 1500 frames, $N=50$ unless stated; replicates in parentheses.}
\label{tab:collective}
\begin{tabular}{lll}
\hline\hline
experiment & group average & self\,+\,social\\
\hline
\parbox[t]{5.0cm}{\raggedright own memory $a=(0.25,0.10)$, local $J_3=0.35$ (40)} & 
\parbox[t]{5.6cm}{\raggedright delay 3 in 22/40, 1 in 13/40; spurious $J_1=0.28$, $J_2=0.12$} & 
\parbox[t]{5.6cm}{\raggedright $(K_s^*,K_c^*)=(2,3)$ in 38/40; $\hat a=(0.245,0.095)$, $\hat J_3=0.348\pm0.024$} \\[1.2ex]

\parbox[t]{5.0cm}{\raggedright own memory of depth 3, no interaction ($N=10$, 30, 100; 40 each)} & 
\parbox[t]{5.6cm}{\raggedright declared in 100\% (exact likelihood: 97--100\%)} & 
\parbox[t]{5.6cm}{\raggedright declared in 1/120; $K_s^*=3$ in 116/120} \\[1.2ex]

\parbox[t]{5.0cm}{\raggedright common drive, no interaction (40)} & 
\parbox[t]{5.6cm}{\raggedright declared in 100\%} & 
\parbox[t]{5.6cm}{\raggedright global 100\%, local 100\%, local with time fields 0\%} \\[1.2ex]

\parbox[t]{5.0cm}{\raggedright common drive, local $J_2=0.3$ (40)} & 
\parbox[t]{5.6cm}{\raggedright declared in 100\%} & 
\parbox[t]{5.6cm}{\raggedright local: $\hat J_2=0.42$; local with time fields: $\hat J_2=0.300\pm0.012$} \\[1.2ex]

\parbox[t]{5.0cm}{\raggedright group-average model true, $J_1=0.25$, $J_4=0.35$ ($N=5$--100; 60 each)} & 
\parbox[t]{5.6cm}{\raggedright $\hat J_4=0.43\pm0.06$ at $N=5$, $0.35\pm0.04$ at $N=100$} & 
\parbox[t]{5.6cm}{\raggedright exact likelihood: $\hat J_4=0.34\pm0.03$ at $N=5$, $0.35\pm0.04$ at $N=100$} \\[1.2ex]

\parbox[t]{5.0cm}{\raggedright own memory, local delay $\Delta=1,2,4,6$ (30 each)} & 
\parbox[t]{5.6cm}{\raggedright delay recovered in 100, 73, 40, 17\%} & 
\parbox[t]{5.6cm}{\raggedright delay recovered in 100\% for every $\Delta$} \\[1.2ex]

\parbox[t]{5.0cm}{\raggedright directions, persistent turning, no interaction (30)} & 
\parbox[t]{5.6cm}{\raggedright declared in 30/30 (delay 1)} & 
\parbox[t]{5.6cm}{\raggedright declared in 0/30} \\[1.2ex]

\parbox[t]{5.0cm}{\raggedright directions, persistence and local $J_3$ (30)} & 
\parbox[t]{5.6cm}{\raggedright delay 3 in 0/30 (always 1)} & 
\parbox[t]{5.6cm}{\raggedright declared in 30/30, delay 3 in 30/30} \\
\hline\hline
\end{tabular}
\end{table*}

\paragraph{Estimation.} The self\,+\,social model~\eqref{eq:selfsocial} is fitted by Newton's method with a sparse Hessian (unit and time fields) and a dense block for the shared couplings. The block of the time fields is diagonal, because each row belongs to one frame, so each Newton step eliminates it exactly (a Schur complement) and solves the remaining dense system for unit fields and couplings, a few hundred parameters; with $6\times10^4$ time fields (baboons at 10\,s) a fit takes seconds, with the same optimum and standard errors as a sparse factorization. Unit and time fields have one flat direction per co-recorded group, $h_i\to h_i+\Delta$, $g_t\to g_t-\Delta$, fixed by a vanishing ridge on the $g_t$; the flat directions are common to all models compared, so BIC differences are unaffected. Self and social depths are selected jointly by BIC on common rows. For directions, $p(\bm n_i^t|\text{past})\propto e^{\bm\theta_i^t\cdot\bm n_i^t}$ is a von Mises law~\cite{MardiaJupp2000} for the unit vector $\bm n_i^t$ of the step, with $\bm\theta_i^t=\mathbf h_i+\mathbf g_t+\sum_\tau a_\tau\bm n_i^{t-\tau}+\sum_\tau J_\tau\bm s_i^{t-\tau}$, the planar spin of Ref.~\cite{Agliari2024}, where $\bm s_i^{t-\tau}$ is the mean unit vector of the neighbors; its negative log-likelihood $\sum[\ln I_0(|\bm\theta|)-\bm\theta\cdot\bm n]$ is convex, and the time-field block is two-by-two block diagonal.

\paragraph{Exactness of the mean-field measure.} The measure $P_A$ of Ref.~\cite{Agliari2024} factorizes over units and frames with one-body weights $e^{x_i^t\theta_t}/Z(\theta_t)$, the first $\tau_M$ frames being fixed by delta functions; evaluated at the data it is $\prod_{t,i}p(x_i^t|\text{past})$ with $\theta_t=H+\sum_\tau J_\tau m^E_{t-\tau}$. Concavity follows as for Eq.~\eqref{eq:loglik}. The stationarity conditions $\sum_{t,i}[x_i^t-\Lambda(\theta_t)]=0$ and $\sum_{t,i}[x_i^t-\Lambda(\theta_t)]m_{t-\tau}=0$ characterize the exact estimator, whereas the original implementation~\cite{Agliari2024} imposes $\Lambda(\theta_t)\simeq m_t$ frame by frame and solves it by least squares after the inversion $G_t=\Lambda^{-1}(m_t)$, a convenient moment approximation.

\paragraph{Aggregation.} Let units be conditionally independent given their own pasts and jointly stationary. For a model of the group average alone, the expected score for $J_\tau$ at $J=0$ is $\E[\sum_i(x_i^t-\bar m)m_{t-\tau}]=N^{-1}\sum_{i,j}\mathrm{Cov}(x_i^t,x_j^{t-\tau})=\bar C(\tau)$ per frame, of order one: $J=0$ is not a solution whenever there is own memory. Linearizing $G_t\simeq\bar\theta+(m_t-\bar m)/\Lambda'(\bar\theta)$, the regression of $G_t$ on lags of $m$ is $1/\Lambda'$ times the autoregression of $m$ on its own lags; since $\mathrm{Cov}(m_t,m_{t-\tau})=N^{-1}\bar C(\tau)$, the factors $N^{-1}$ cancel in the Yule--Walker equations, which gives Eq.~\eqref{eq:aggregation}. For the self\,+\,social model with leave-one-out regressors, the tower property gives $\E[(x_i^t-\E[x_i^t|x_i^{<t}])\,m^{(-i)}_{t-\tau}]=0$, because $m^{(-i)}$ is independent of unit $i$: the social score vanishes at the truth.

\paragraph{Common drives.} A shared stationary field $u_t$ with autocovariance $C_u$ induces $\mathrm{Cov}(x_i^t,x_j^{t-\tau})\simeq\Lambda'^2C_u(\tau)$ between non-interacting units, to linear order. With unit and time fields, the identity $m^{(-i)}_{t-\tau}=\frac{N_t}{N_t-1}m_{t-\tau}-\frac1{N_t-1}x_i^{t-\tau}$ ($N_t$ units in the frame) shows that the global regressor is a function of the frame, absorbed by $g_t$, plus a multiple of the own lag. For $\tau\le K_s$ a global coupling is not identifiable (exactly so when $N_t$ is constant); for $\tau>K_s$ it estimates $-(N_t-1)$ times the omitted own memory at that lag. A local regressor is not in the span of the time fields and own lags unless all neighborhood means coincide. On synthetic data [\figref{fig:collective}(a)], every reading except local coupling with time fields declares a coupling under a common drive alone, and only that reading recovers a planted local coupling without bias.

\paragraph{Mean-field theory.} By the law of large numbers, $m_t\to\Lambda(H+\sum_\tau J_\tau m_{t-\tau})$. Linearizing around a fixed point gives the characteristic equation $1=\Lambda'(\theta^*)\sum_\tau J_\tau z^{-\tau}$. For non-negative couplings, $f(z)=\Lambda'\sum_\tau J_\tau z^{-\tau}$ is decreasing on $z>0$ and $|f(z)|\le f(|z|)$, so a root with $|z|\ge1$ exists if and only if $f(1)=\Lambda'J_{\rm tot}\ge1$; when the delays have no common divisor the critical root is $z=1$ alone (a fold). Since $\Lambda'\le L^2/12$, with equality at $\theta=0$, bistability requires $J_{\rm tot}>12/L^2$, the cusp being at $\tilde\theta=0$, $m=L/2$. For a single delay $d$ the map decouples into $d$ sub-chains $y^{(r)}_n=m_{nd+r}$ obeying $y_{n+1}=\Lambda(H+Jy_n)$; for $J<0$ each undergoes a flip at $\Lambda'|J|=1$, and a lag-1 term $a_1>0$ in the characteristic equation $z^d=\Lambda'J+\Lambda'a_1z^{d-1}$ favors the in-phase mode, of period $2d$. For finite $N$ and one lag, $\delta m_t=\Lambda'J\delta m_{t-1}+\eta_t$ with $\mathrm{Var}\,\eta=\Lambda'/N$ gives $N\,\mathrm{Var}(m)=\Lambda'/[1-(\Lambda'J)^2]$ [\figref{fig:collective}(b)].

\paragraph{Own memory in mean field.} With lag-1 self coupling $a\ge0$, propagation of chaos makes each unit, at a stationary state $m^*$, an independent chain with field $h_{\rm eff}=H+J_{\rm tot}m^*$ and stationary law $\pi(x)\propto e^{h_{\rm eff}x}Z(h_{\rm eff}+ax)$, Eq.~\eqref{eq:stationary}. The self-consistency is $m^*=\Psi(H+J_{\rm tot}m^*)$, with $\Psi(h)=\E_\pi[x]$, and the symmetry line is $H+(J_{\rm tot}+a)L/2=0$. A field perturbation propagates as $\delta m_t=\sum_{s\ge0}R(s)\,\delta h_{t-s}$, with $R(s)=\mathrm{Cov}(x_s,\epsilon_0)$ the response of one unit, $\epsilon_0=x_0-\E[x_0|\text{past}]$ being the innovation; for $a\ge0$, $R(s)\ge0$, the argument above applies to $1=\hat R(z)\sum_\tau J_\tau z^{-\tau}$, and stability is lost at $J_{\rm tot}\chi=1$ with $\chi=\sum_sR(s)=\partial\Psi/\partial h$. To leading order in $1/N$, $N\,\mathrm{Var}(m)=\int\frac{d\omega}{2\pi}S_x(\omega)|1-\hat J(\omega)\hat R(\omega)|^{-2}$, with $S_x$ the spectrum of one unit. We compute $R$, $S_x$ and $\chi$ from the transition matrix of the chain on 400 Gauss--Legendre nodes.

\paragraph{Synthetic collectives.} The speed-Vicsek model places $N$ units in a periodic box; headings follow Vicsek alignment~\cite{Vicsek1995}, and log-speeds are drawn from Eq.~\eqref{eq:selfsocial} with planted own memory, local (three nearest neighbors) or global couplings and, when stated, a common Ornstein--Uhlenbeck drive; steps are then recomputed from the positions, as in an experiment. Individual fields are drawn from $U(-1.3,-0.5)$. Table~\ref{tab:collective} collects the experiments. Directions behave like speeds: for units with persistent turning and no interaction, the group-average analysis of directions declares delayed interactions in every replicate, the self\,+\,social von Mises model in none. With AIC instead of BIC, the self\,+\,social model itself declares spurious couplings in 27\% of interaction-free data sets, the value of Eq.~\eqref{eq:woodroofe} again.

\paragraph{Time-shift null.} In a time-shift surrogate every unit's track is shifted by an independent random circular offset, which keeps its own memory, heterogeneity and marginal law and destroys every synchrony between units, interactions and common drives alike; a permutation of frames destroys own memory as well, and therefore tests memory and synchrony together.

\paragraph{\rev{Time fields in small groups.}} \rev{A time field is estimated from the $N_t$ units present in its frame. Since the number of fields grows with the number of frames while $N_t$ stays fixed, this is an incidental-parameter problem~\cite{NeymanScott1948}, and the shared couplings need not converge to their true values. We measured the effect on synthetic groups in which every unit is coupled to one other unit, drawn at random in each frame, with unit fields and a common drive. For log-speeds it is negligible: with a self coupling 0.4 and a social coupling 0.3, the fits with time fields return social couplings 0.32, 0.32, 0.30 and 0.30 for 3, 5, 8 and 26 units, against 0.30, 0.31, 0.30 and 0.30 without them. For headings it is not, because in the von Mises law a coupling is also a concentration, and a field per frame absorbs part of the angular spread, just as a mean direction estimated from a few samples underestimates the dispersion about it. All couplings are then inflated by a common factor. For weakly concentrated headings ($a_1=0.7$, $J_1=0.45$) the factor is close to $N_t/(N_t-1)$: 1.56, 1.29, 1.16 and 1.05 for 3, 5, 8 and 26 units. For headings as concentrated as those of the pigeons [$a_\tau=(185,-82,21)$, $J_\tau=(17.6,-11.0,3.8)$] it is 1.9, 1.5 and 1.3 for 3, 5 and 8 units, the ratio of social to self coupling being preserved within 8\%, and it disappears without time fields. A null coupling acquires at most a small positive offset, between 1 and 2.5 for three units and below 1 for five, against self couplings of order $10^2$, which the time-shifted surrogates of the same groups carry as well. Signs and ratios of heading couplings, and their comparison with time-shifted surrogates, are therefore unaffected, whereas their absolute values in small groups are upper bounds.}

\paragraph{Neighbors in real groups.} The local regressor of unit $i$ at delay $\tau$ is the mean of the variable over the units of distance ranks $r_0+1,\dots,r_1$ from $i$ at the emission frame $t-\tau$: the three nearest for baboons, and the nearest one for pigeons, whose flocks have three to eight birds in a frame, so that a larger neighborhood would be the whole flock and, as shown above, not identifiable with time fields. The decay with distance is measured by single-neighbor regressors of ranks 1, 2, 3, 6 and 12 (baboons) or 1, 2 and 3 (pigeons), fitted jointly with three delays each; the asymmetry, by splitting the six (baboons) or three (pigeons) nearest neighbors into those ahead of and those behind $i$ along the group heading, the direction of the mean step at $t-\tau$. Anchoring the neighborhood at the current position of $i$ instead (the units that at $t-\tau$ were nearest to where $i$ is now) also picks up fields fixed in space, such as trails, which time fields cannot absorb. In the time-shift null the positions are shifted with the steps, so the neighbors of a shifted unit are whoever passed near the same place at another time: the null removes synchrony but keeps fields fixed in space, and its coupling to the nearest neighbor need not vanish.

\bibliography{refs}

\end{document}